%% file: main_mark.tex
\documentclass[aip,amsmath,amssymb,preprint,author-year]{revtex4-1}

\input{inputs_mark.tex}

\begin{document}

\title{A Statistical Study of the Compressible Energy Cascade Rate in Solar Wind Turbulence: Parker Solar Probe Observations} %Title of paper

\author{M. Brodiano}
\email[Email: ]{mbrodiano@df.uba.ar}
\affiliation{Instituto de F\'isica de Buenos Aires, CONICET-UBA, Ciudad Universitaria, 1428 Buenos Aires, Argentina}
\affiliation{Departamento de F\'isica, Facultad de Ciencias Exactas y Naturales, Universidad de Buenos Aires, Ciudad Universitaria, 1428, Buenos Aires, Argentina.}
\author{P. Dmitruk}
\affiliation{Instituto de F\'isica de Buenos Aires, CONICET-UBA, Ciudad Universitaria, 1428 Buenos Aires, Argentina}
\affiliation{Departamento de F\'isica, Facultad de Ciencias Exactas y Naturales, Universidad de Buenos Aires, Ciudad Universitaria, 1428, Buenos Aires, Argentina.}
\author{N. Andr\'es}
\affiliation{Departamento de F\'isica, Facultad de Ciencias Exactas y Naturales, Universidad de Buenos Aires, Ciudad Universitaria, 1428, Buenos Aires, Argentina.}
\affiliation{Instituto de Astronomía y Física del Espacio, CONICET-UBA, Ciudad Universitaria, 1428, Buenos Aires,Argentina}

%email corresponding author: \href{mbrodiano@df.uba.ar}{mbrodiano@df.uba.ar}

\date{\today}
\begin{abstract} 
We investigated incompressible and compressible magnetohydrodynamic (MHD) energy cascade rates in the solar wind at different heliocentric distances. We used in situ magnetic field and plasma observations provided by the Parker Solar Probe (PSP) mission and exact relations in fully developed turbulence. To estimate the compressible cascade rate, we applied two recent exact relations for compressible isothermal and polytropic MHD turbulence, respectively. Our observational results show a clear increase of the \ADD{absolute value of the} compressible and incompressible cascade rates as we get closer to the Sun. Moreover, we obtained an increase in both isothermal and polytropic cascade rates with respect to the incompressible case as compressibility increases in the plasma. Further discussion about the relation between the compressibility and the heliocentric distance is carried out. \ADD{Furthermore}, we compared both exact relations as compressibility increases in the solar wind and although we note a slightly trend to observe larger cascades using a polytropic closure, we obtained essentially the same cascade rate in the range of compressibility observed. \ADD{Finally, we investigated the signed incompressible and compressible energy cascade rates and its connection with the real cascade rate.}
\end{abstract}


\maketitle %\maketitle must follow title, authors, abstract and \pacs

\section{Introduction}\label{sec:intro}

The solar wind is a well-studied supersonic plasma flow, from the Sun's atmosphere to the edge of the heliosphere, characterized by a turbulent energy cascade rate \citep{Tu1995,Fra2019}. The availability of in situ measurements from various orbiting spacecraft through different heliocentric distances allow a deep understanding of the essential role of turbulence from the large scales up to the kinetic scales in the solar wind plasma \citep{A2013,bruno2013,K2015,Ch2016}. In this sense, the Parker Solar Probe \citep{Fo2016} (PSP) mission has been exploring the inner heliosphere since 2018, approaching to the Sun with each orbit and enabling us to study the evolution of turbulence and to compare observations with theoretical predictions.

%Therefore, the solar wind provides a natural laboratory for the study of astrophysical plasma turbulence \citep{S2020,Ch2020,H2020a,A2019b,A2021,A2022,H2019,Ha2017b,K2015,A2013,H2012,O2011}. 

A prevailing challenge in the solar wind community is to enhance the models of the turbulent heating of the plasma, with particular focus in the near-Earth space \citep{S2020,Ha2017b} and the magnetosphere environments \citep{Ch2020,H2020a,A2019b,A2021,A2022,H2019,K2015,A2013,H2012,O2011}. Observations have shown that the solar wind proton temperature tends to decrease as a function of the radial distance from the Sun more gradually than the prediction of the adiabatic expansion model of the solar wind \citep{M1982c,V2007,Pi2020}. Although several scenarios have been proposed to explain these observations, the main candidate is certainly the local heating of the solar wind plasma via the turbulent cascade \ADD{\citep{bruno2013,M2011,Ma2023}}. Therefore, the energy at the largest scales in the solar wind will cascade within the inertial range until it reaches the ion scales where it is eventually transformed into thermal or kinetic energy of the plasma particles (see \citet{S2020,K2015}).

%The exact relations relate the longitudinal structure function of the turbulent variables (the plasma density or velocity field or magnetic field, in the simplest cases) taken in two points to the spatial increment $\ell$ that separates them.

%Under the assumption of homogeneity and isotropy, the so-called $4/5$ law \citep{K1941a,K1941b} predicts a linear scaling for the longitudinal third-order structure function of the velocity field with the distance between points (only valid in the inertial range).

Several theoretical efforts have been made in order to provide an estimation of the turbulent energy cascade rate at the solar wind. The first exact relation use the von-Kármán-Howarth dynamical equation \citep{vkh1938}. This exact relation gives an expression for the energy dissipation or cascade rate $\varepsilon$ as a function of structure functions of the velocity turbulent field \citep[e.g.,][]{MY1975,F1995}. \citet{Ga2011} and \citet{B2014} generalized this exact relation to compressible isothermal and polytropic hydrodynamic (HD) turbulence, respectively. \citet{Ga2011} has found a new term on the inertial range that behaves similarly as a source or a sink for the mean energy cascade rate. In particular, this new term is a product of increments and local mean values of the plasma variables, on the contrary to the typical structure functions present in the incompressible flows.

%The authors showed that the introduction of a uniform magnetic field simplifies significantly the exact relation for which a simple phenomenology may be given. 

In the case of magnetized plasmas, the first exact relation was derived by \citet{P1998b,P1998a} for homogeneous, isotropic and incompressible magnetohydrodynamics (IMHD) turbulence. The authors recovered a scaling law for mixed third-order longitudinal structure and correlation functions. This exact law has been the subject of several numerical tests \citep[e.g.,][]{Sorriso2002,Mi2009,Bo2009,W2010,Verdini_2015}; it has been used for the estimation of the incompressible cascade rate in space plasmas \citep{So2007,Sa2008,Co2015,Si2008,Ma2008} and the magnetic and kinetic Reynolds numbers \citep{WEY2007} in solar wind turbulence, and for the large-scale modeling of the solar wind \citep{M1999,Mc2008,MacBride2005}. \citet{B2013} derived an exact relation for some two-point correlation functions of the fields for isothermal compressible magnetohydrodynamic (CMHD) turbulence, written in terms of the flux or the source terms. On the other hand, \citet{A2017b} revisited the latter work and expressed the exact law as a function of a more adequate set of plasma variables, i.e.,  the plasma mass density, the plasma velocity field and the compressible Alfvén velocity. This theoretical work presented four different types of terms that are involved in the non-linear cascade of energy in the inertial range: the hybrid terms (which can be written either as flux or source terms), the $\beta$-dependent terms and the well-known flux and source terms (see, \citet{A2019,F2020}). Recently, \citet{S2021} proposed a more general method that allows to derive the exact relation for any turbulent isentropic flow (i.e., constant entropy). The authors demonstrated that the well-studied MHD exact laws (incompressible and isothermal) and the new (polytropic) one can be obtained as specific cases of the general exact relation when the corresponding closure is specified. These formulations for IMHD and CMHD turbulence are used in the present study to estimate the non-linear transfer of energy in the solar wind.

A first attempt to include the compressibility in estimating the energy cascade rate using in situ observations was reported by \citet{C2009b} using a phenomenological model and Ulysses observations. The authors found a significant increase in the turbulent cascade rate with respect to the incompressible exact law. However, those results were based on a heuristic model, using pseudo-energies (i.e., fluxes through the scale of the increments of the Elsässer fields), which are not conserved in CMHD theory \citep{M1987}. In the same framework, \citet{M2008} showed that the compressible turbulent cascade seems to be able to supply the energy needed to account for the local heating of the non‐adiabatic solar wind. \citet{B2016c} and \citet{H2017a} have studied the impact of the compressible fluctuations in the energy cascade in the solar wind using a reduced form of the exact relation for CMHD turbulence and in situ observations from the Time History of Events and Macroscale Interactions during Substorms (THEMIS) \citep{Au2009} spacecraft. The authors found that the compressible fluctuations are shown to amplify by several order of magnitude the turbulent cascade rate with respect to the incompressible model. Recently, \citet{B2020} have computed the incompressible energy transfer rate between 35 and 55 solar radius using PSP observations during the first encounter. \citet{A2021} extended previous observational studies computing the compressible energy transfer rate from $\sim 0.2$ au up to $\sim 1.7$ au, using PSP, THEMIS and Mars Atmosphere and Volatile EvolutioN (MAVEN) observations. The authors showed that, depending on the level of compressibility in the plasma, the different terms in the compressible exact relation were shown to have different impact in the total cascade rate (where the incompressible terms are included). Moreover, using more than 2 years of PSP observations, \citet{A2022} studied the incompressible energy cascade rate using isotropic and anisotropic exact relations. The authors found a connection between the heliocentric distance, the local temperature of the plasma and the energy cascade rate, with a clear dominance of the perpendicular cascades over the parallel cascades as PSP approaches the Sun.

%\ADD{In particular, the authors confirmed that density fluctuations in the solar wind amplify the cascade rate with respect to the incompressible model and increase the compressible component of the exact law becoming of the order of the Yaglom-like component in the inertial range. In addition, their observational results show a correlation between the solar wind temperature and the compressible energy cascade rate, showing an increase in the temperature which is connected with fast solar winds speeds.}

In the present paper, we use the PSP observations (magnetic field and plasma moments) at different heliocentric distances to compute the \ADD{absolute value of the} compressible and incompressible energy cascade rates. \ADD{We investigated as well the signed incompressible and compressible energy cascade rates and its connection with the real cascade rate.} Our goal is to discuss the impact of the plasma compressibility and heliocentric distance in the energy cascade in the solar wind. The paper is organized as follows: in section \ref{sec:theo} we recall the theoretical CMHD set of equations and present briefly the main steps to derive the exact law for fully developed CMHD turbulence. In section \ref{sec:Selec}, we describe the data set composed of more than 3000 PSP events and the selection criteria used in the present study. In section \ref{sec:Res}, we present the main observational results. Finally, in section \ref{sec:Discussion} we provide a summary and discussion of our main findings. 

\section{Theoretical Models and Exact Relations}\label{sec:theo}

\subsection{Compressible MHD model}\label{sec:model}

In the present work, we used the 3D CMHD model given by the mass continuity equation, the momentum equation for the velocity field $\mathbf{u}$, the induction equation for the magnetic field $\mathbf{B}$ and the differential Gauss's law (solenoidal condition $\boldsymbol\nabla\cdot{\bf B}=0$). These equations can be written as \citep{M1987},
\begin{align}\label{continuidad}
    \frac{\partial \rho}{\partial t}&=-\boldsymbol\nabla\cdot(\rho {\bf u}), \\ \label{NS}
     \frac{\partial {\bf u}}{\partial t}&=-{\bf u}\cdot\boldsymbol\nabla{\bf u}+{\bf u}_A\cdot\boldsymbol\nabla {\bf u}_A -\frac{{\boldsymbol{\nabla}}  (P+P_M)}{\rho}-{\bf u}_A(\boldsymbol\nabla\cdot{\bf u}_A)+\mathbf{f}_k+{\bf d}_k, \\ \label{induccion}
    \frac{\partial {\bf u}_A}{\partial t} &=-{\bf u}\cdot\boldsymbol\nabla{\bf u}_A+{\bf u}_A\cdot\boldsymbol\nabla {\bf u}-\frac{{\bf u}_A}{2}(\boldsymbol\nabla\cdot{\bf u})+{\bf f}_m+{\bf d}_m, \\ \label{Gauss}
    0 &= {\bf u}_A\cdot\boldsymbol\nabla\rho+2\rho(\boldsymbol\nabla\cdot{\bf u}_A),
\end{align}
where ${\bf u}$ is the velocity field (assuming a zeroth background flow speed) and ${\bf u}_A={\bf B}/\sqrt{4\pi\rho}$ is the compressible Alfvén velocity with $\mathbf{B}$ the total magnetic field and $\rho$ the mass density. Note that, the time dependence enters through ${\bf u},~{\bf B}$ and $\rho$. In addition, $P$ is the scalar isotropic pressure and $P_M = \rho u_A^2/2$ is the magnetic pressure. In the present work, we use two different equations of state which allow us to close the hierarchy of the fluid equation (so there is no need of an energy equation): the isothermal case $P=C^2_{s} \rho$, where $C_{s}$ is the (constant) sound speed and the polytropic case $\gamma P=C_s\rho$, where $\gamma=5/3$ is the ratio of the heat capacity at constant pressure to heat capacity at constant volume and $C_s$ is the (variable) sound speed. For a detailed discussion about the two equation of state see Section 2.1 in \citet{S2021}. Finally, ${\bf f}_{k,m}$ are the respectively mechanical and the curl of the electromotive large-scale forcing, and ${\bf d}_{k,m}$ are respectively the small-scale kinetic and magnetic dissipation terms.

\subsection{Exact relation in CMHD turbulence}\label{sec:laws}

In the CMHD model, the density total energy $E({\bf x})$ and the density-weighted cross helicity $H({\bf x})$ are given by,
\begin{align}
    E({\bf x})&=\frac{\rho}{2}({\bf u}\cdot{\bf u}+{\bf u}_A\cdot{\bf u}_A)+\rho e , \\
    H({\bf x})&= \rho~({\bf u}\cdot{\bf u}_A),
\end{align}
where $e$ is the internal energy. It is worth mentioning that a polytropic process is a thermodynamic process that obeys the relation: $PV^n=C$, where $P$ is the pressure, $V$ is a control volume, $n$ is the polytropic index, and $C$ is a constant. The polytropic process equation describes expansion and compression processes which include heat transfer. Some specific values of $\gamma$ correspond to particular cases like, $n = 0$ for an isobaric process and $n=\infty$ for an isochoric process. In addition, when the ideal gas law applies, $n = 1$ for an isothermal process and $n = \gamma$ for an isentropic process. Since these definitions are compatible with previous studies in the literature \citep[see,][]{Ga2011, B2013, B2014, B2016c, H2017a, Ha2017b, A2019b, A2020, S2021}, we decided to use the names {\it isothermal} for a plasma that obey $P_\text{iso} \sim \rho$, and consequently $e_\text{iso} = C^{2}_{s0}\ln(\rho/\rho_0)$, and {\it polytropic} for a plasma with $P_\text{pol} \sim \rho^\gamma$ and $e_\text{pol} = (C_s^2-C^2_{s0})/(\gamma(\gamma-1))$. In both cases, the sound speed $C_s$ (and its mean value $C_{s0}^2$) can be obtained from the perfect gas equation, $C_s^2 = \gamma k_B T_p/m_p$, where $T_p$ and $m_p$ are the proton temperature and mass, with $n = 1$ and $n = \gamma = 5/3$ for the isothermal and polytropic case, respectively \citep[see,][]{Ga2011,B2013}. Finally, it is worth mentioning that any conclusion in the present paper is limited by these assumptions about the thermodynamics discussed here.

While the total energy is one of the ideal invariants, the density-weighted cross helicity is not. Nevertheless, both quantities are essential for the derivation of the exact law in CMHD turbulence \citep[see,][]{A2017b,S2021}. In particular, we can define the two-point correlation function associated with the total energy, the helicity and the magnetic field by,
\begin{align}\label{RE}
    R_E({\bf x},{\bf x}')&=\frac{\rho}{2}({\bf u}\cdot{\bf u}'+{\bf u}_A\cdot{\bf u}'_A)+\rho e', \\ \label{RH}
    R_H({\bf x},{\bf x}')&=\frac{\rho}{2}({\bf u}\cdot{\bf u}'_A+{\bf u}_A\cdot{\bf u}'), \\ \label{RB}
    R_B({\bf x},{\bf x}')&=\frac{\rho}{2}({\bf u}_A\cdot{\bf u}'_A)
\end{align}
where the prime denotes field evaluation at ${\bf x}'={\bf x}+\boldsymbol\ell$ ($\boldsymbol\ell$ being the displacement vector). Using Eq. \eqref{continuidad}-\eqref{Gauss}, the expressions \eqref{RE}-\eqref{RB} and following the usual assumptions for fully developed homogeneous turbulence (i.e., infinite kinetic and magnetic Reynolds numbers and a steady state with a balance between forcing and dissipation) \citep[see,][]{B2018,Ba2020}, an exact relation for compressible MHD turbulence can be obtained in the inertial range as,
\begin{equation}
    -2\varepsilon_C = \frac{1}{2}\boldsymbol\nabla_\ell\cdot{\bf F}_C+S_C+S_H+M_{\beta},
    \label{TermComplete}
\end{equation}
where $\varepsilon_C$ is the total compressible energy cascade rate, ${\bf F}_C$ is the total compressible flux (defined below)  and $S_C$, $S_H$ and $M_{\beta}$ are the so-called source, hybrid and $\beta$-dependent terms, respectively. For a detailed derivation and the explicit expressions of the total compressible energy cascade rate, see \citet{A2017b,S2021}. It is worth mentioning that, the derivation of the exact law \eqref{TermComplete} does not require the assumption of isotropy \citep{A2022} and that it is independent of the dissipation mechanisms acting in the plasma (assuming that the dissipation acts only at the smallest scales in the system). In the present paper, in order to estimate the compressible energy cascade rate, we shall consider only the flux terms. The main reason is that the source, hybrid and $\beta$-dependent terms require computing the divergence of the plasma and the compressible Alfvén speeds, which can be done only using multi-spacecraft observations. Moreover, numerical results for supersonic and subsonic HD and MHD turbulence have shown that the source, hybrid and $\beta$-dependent terms are negligible with respect to the flux term in the inertial range \citep[e.g.,][]{Kr2013,A2018,A2019b,F2020}. 

The total compressible flux is a combination of two terms of different nature, a Yaglom-like term,
\begin{equation}\label{yaglom}
    {\bf F}_{1C} = \langle[\delta (\rho {\bf u})\cdot\delta{\bf u}+\delta (\rho {\bf u}_A)\cdot\delta{\bf u}_A]\delta{\bf u}-[\delta (\rho {\bf u})\cdot\delta{\bf u}_A+\delta{\bf u}\cdot\delta (\rho {\bf u}_A)]\delta{\bf u}_A\rangle,
\end{equation}
which is the compressible generalization of the incompressible term \citep{P1998a,P1998b}, and a purely compressible flux term,
\begin{equation}\label{com_flux}
    \mathbf{{F}}_{2C} = 2\langle\delta\rho\delta e \delta {\bf u}\rangle,
\end{equation}
which is a contribution to the energy cascade rate due to the presence of density fluctuations in the plasma \citep{A2017b,S2021}. Here, we have introduced the usual increments, i.e., $\delta \alpha =  \alpha' - \alpha$ (with $\alpha$ any scalar or vector function) and the angular bracket $\langle \cdot\rangle$ denotes an ensemble average. It is worth mentioning that the difference between the two compressible models is in the form of the specific internal energy $e$, therefore, the polytropic and isothermal cascade rate differences would arise only from the compressible flux term \eqref{com_flux}. Finally, assuming statistical isotropy, we can integrate the flux terms in \eqref{TermComplete} over a sphere of radius $\ell$ to obtain a scalar relation for isotropic turbulence. In compact form, \eqref{TermComplete} can be written as,
\begin{equation}
    -\frac{4}{3} \varepsilon_C \ell = F_{1C}+F_{2C},
    \label{Compre}
\end{equation}
where $F_{1C}+F_{2C} = ({\bf F}_{1C}+{\bf F}_{2C})\cdot\hat{{\bf u}}_{sw}$ is the flux term projected into the mean plasma flow velocity field ${\bf u}_{sw}$. Note that, we called $\varepsilon_{1C}+\varepsilon_{2C} = (-3/4\ell)(F_{1C}+F_{2C})$.

In the incompressible limit $\rho\to\rho_0$, we recover the Politano and Pouquet law for fully developed incompressible MHD turbulence, 
\begin{equation}
      -\frac{4}{3} \varepsilon_I \ell = F_{I},
      \label{Incom}
\end{equation}
where $F_I$ is the projection of ${\bf F}_I = \rho_0\langle[(\delta{\bf u}^2)+(\delta{\bf u}_A^2)]\delta{\bf u}-2(\delta{\bf u}\cdot\delta{\bf u}_A)\delta{\bf u}_A\rangle$ along the mean plasma flow velocity field \citep{P1998b,P1998a}. Assuming the Taylor hypothesis, i.e., $\ell \approx u_0\tau$, where $\tau$ is the time lag and $u_0$ is obtained by averaging over each time interval, all the quantities that are indexed by 0, Eqs. \eqref{Compre} and \eqref{Incom} can be expressed as a function of the time lags $\tau$.

% Estoy aca NA

\section{PSP Observations and Data Selection Criteria}\label{sec:Selec}

In order to estimate the compressible energy cascade rate, we employed magnetic field measurements by the FIELDS flux gate magnetometer (MAG), along with proton density, velocity and temperature data from the Solar Probe Cup (SPC) of the SWEAP instrument suite \citep{Ba2016,Fo2016,K2016,B2019,Ka2019,C2020}. Our analysis on the PSP observations involved a time interval between November 2, 2018 (Encounter 1), and December 30, 2020 (Encounter 6). This data set was divided into a series of samples of equal duration of 60 minutes. This particular time duration ensures having at least one correlation time of the turbulent fluctuations for each particular heliocentric distance \citep{P2020,H2017a}. To generate uniform time series, we re-sampled all the variables to 30 s time resolution. Thus, there are 10944 events using this data set. Regarding the presence of switchbacks \ADD{\citep{Bour2020,Dudok2020,He2021}}, the well-known rapid polarity reversals in magnetic fields, we did not take it into account in the present paper. To do this, we used the following criteria so as to discard the presence of switchbacks (keeping the intervals where the energy cascade rate is approximately constant).

\begin{figure}
    \centering
    \includegraphics[width=0.6\linewidth]{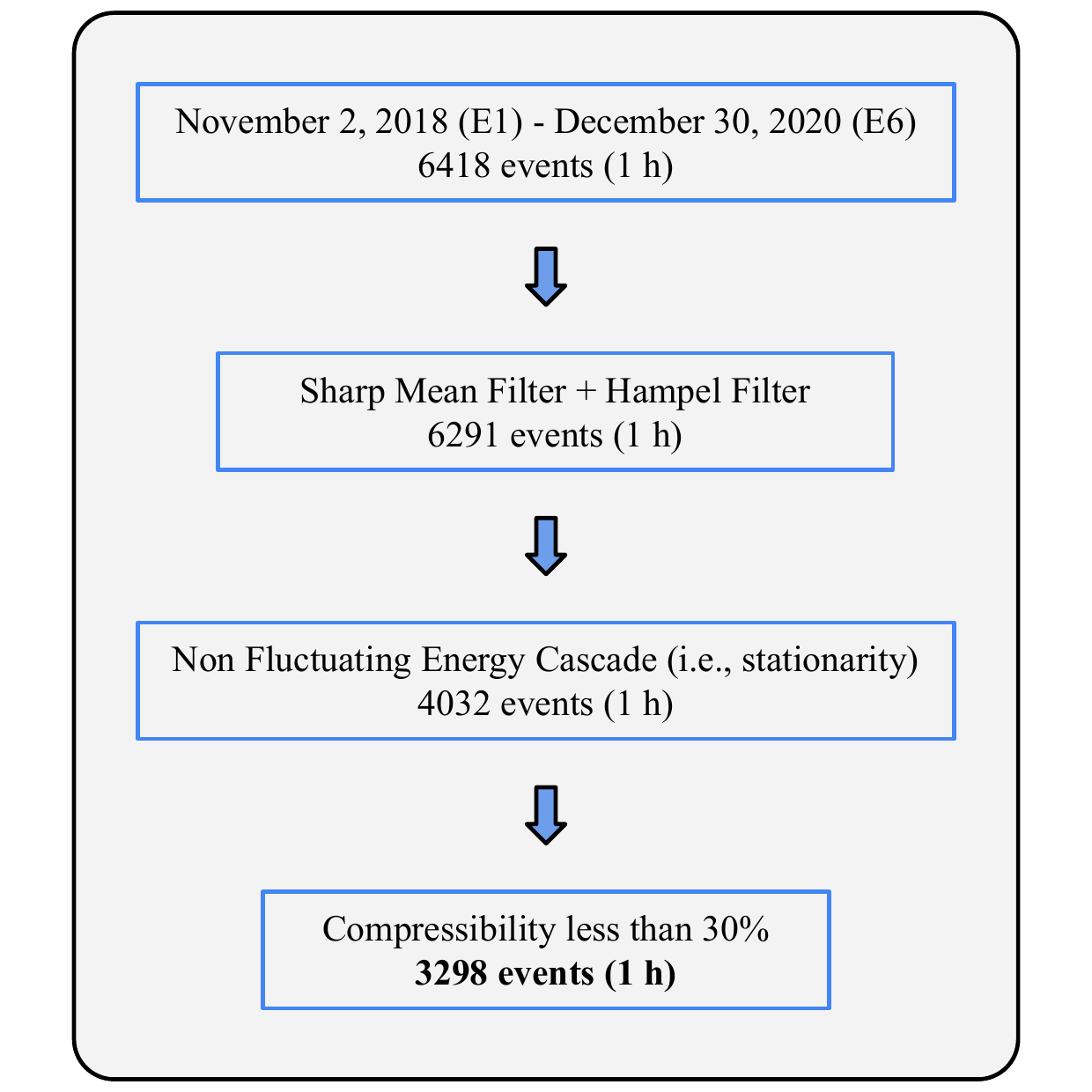}
    \caption{Schematic diagram of the filters used in order to remove spurious spikes from the data set.}
    \label{Esquema}
\end{figure}

In our data set, spurious spikes in the SPC moments, which are remnants of poor quality of fits, were removed. To accomplish this, consecutive filters were applied, as Figure \ref{Esquema} shows. First, for each event we applied a {\it sharp} mean value filter, which detect each spurious data using a threshold in the time series and replace it with the average between the previous and the next valid observation. When we say valid observation, we are referring to a real value of the data set, i.e., values that have passed the first threshold that we applied. Although this filter removes most of the spurious data, in some cases leaves a fictitious large data value. Therefore, we applied the well-known Hampel filter in order to detect this fictitious outliers \citep{Davies1993,LIU2004,Pearson2016}. In few words, the Hampel filter uses a moving window implementation of predetermined size (in particular, we used a window of 8 points) to compute the local median $m_i$ and the local standard deviation $s_i$. Then, for each point $x_i$,
\begin{equation}
    y_i =
    \begin{cases}
    x_i & \text{if $|x_i-m_i| \le n_{\sigma}S_i$}, \\
    m_i & \text{if $|x_i-m_i| > n_{\sigma}S_i$},
    \end{cases}
\end{equation}
if the absolute difference of the value of that point and the local median, $|x_i-m_i|$ is above a threshold defined as $n_{\sigma}$ times the local standard deviation, the value is replaced by the median. If not, the algorithm leaves the current point unchanged and proceeds to the next point. We applied this technique using a local window of 8 points and choosing a threshold value of $n_{\sigma} = 3$ \citep{riddhi2018}. This choice of parameters help eliminate most of the undetected spikes and smooth the data set probably altered by the previous filter.

\begin{figure}[b]
    \centering
    \includegraphics[width=0.85\linewidth]{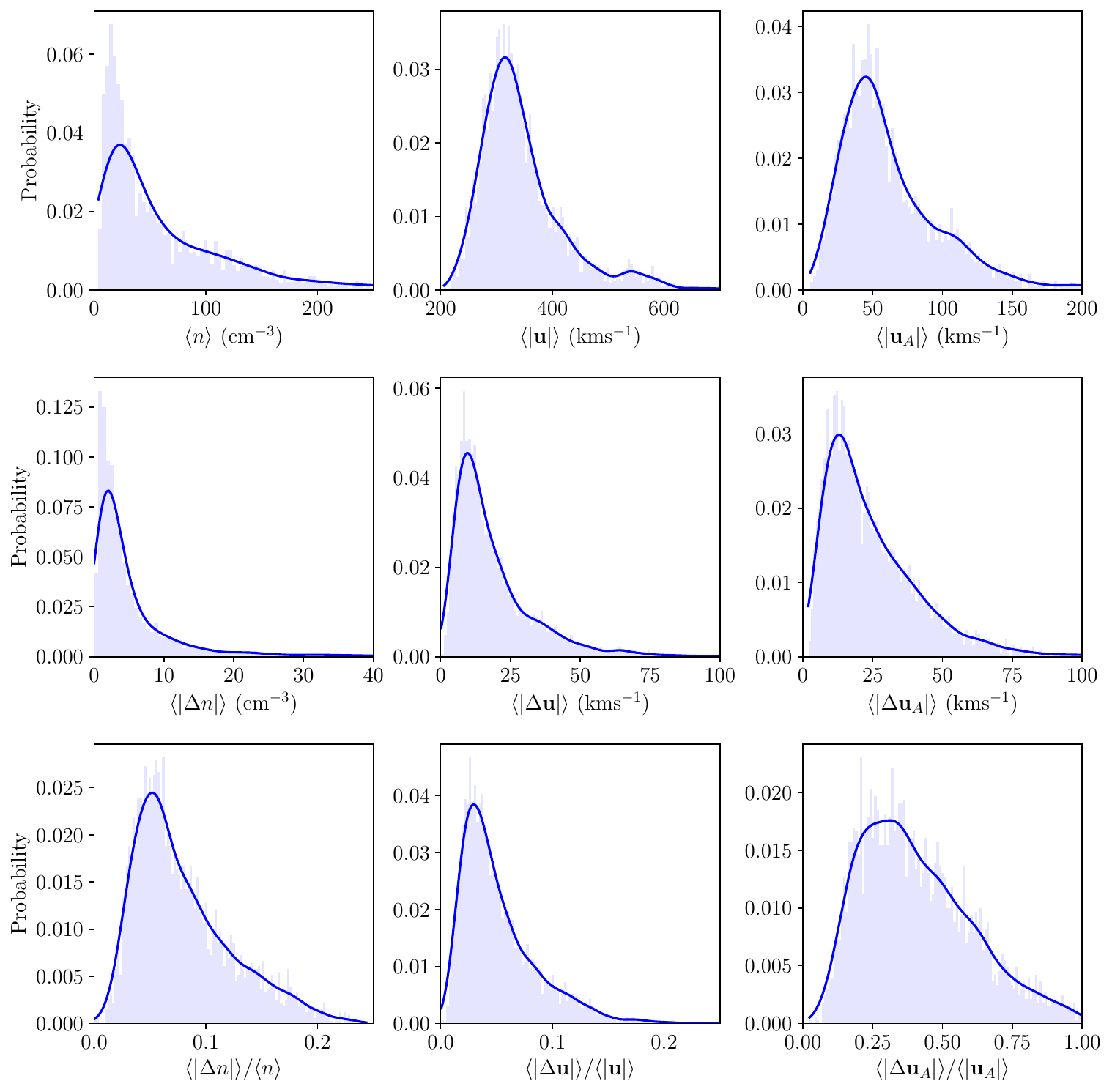}
    \caption{The occurrence rate for the proton density, the proton and Alfvén velocity absolute mean values,in the first row, the corresponding fluctuations in the second row and the ratio between these magnitudes in the third row, respectively.}
    \label{Histogram}
\end{figure}

In addition, we considered only intervals that did not show large fluctuations of the energy cascade rate over the large MHD scales, typically we retained events with $\text{std}(\varepsilon_I)/\text{mean}(|\varepsilon_I|)<0.85$, where we have computed the standard deviation and the mean of the absolute value of the energy cascade over each event of 1 hour duration in the MHD scales. Finally, we analyzed the distribution of density fluctuations of the filtered data set. Figure \ref{Histogram} shows the occurrence rate for all the analyzed events for the numerical density, velocity, and Alfvén velocity field absolute values and their fluctuations, respectively. We can relate the distribution of density fluctuations with the compressibility ratio defined as $\sqrt{\langle\rho^2\rangle-\langle\rho\rangle^2}/\langle\rho\rangle$. Since most of the events did not show high levels of compressibility rate, we only kept the events whose compressibility reached up to 30$\%$. This leaves us a data set of 3298 events.
\section{Observational Results}\label{sec:Res}

In order to compute the compressible and incompressible energy cascade rate, we constructed correlation functions of the different turbulent fields at different time lags $\tau$ in the interval $\tau=[30,3480]$s. Therefore, once we had the energy cascade as a function of the time increments and in order to quantify statistically these increments, we took the average of the absolute values in the largest MHD scales, i.e., $\tau=[1000,3000]$s to obtain representative values of each event in the large scales \citep{H2017a,A2020}.  It is worth to note that, the sign of the cascade rate can vary within the inertial range \citep{So2007} (showing a constant negative or positive sign for different time lags in the MHD range), which is related to the direct and inverse nature of the energy cascade rate \citep[see,][]{Co2015,H2017a,A2020}. \ADD{In particular, in the present paper, we focused on the study of the absolute value of the energy cascade in the MHD scales. In section \ref{Sign}, we have also included an analysis of the limitations when using the absolute value of the cascade by making a comparison with the signed cascade rate.} Finally, for the statistical convergence of the energy cascade rate, we analysed in detailed the correlation functions to assure its convergence, i.e. the constant value in the MHD scales). Further discussion is described in the Appendix \ref{appe}. 

%we have studied the absolute value of the energy cascade, with a particular focus on the MHD scales. Therefore, there is a clear limitation of analyzing the unsigned energy cascade rates. The sign of the energy cascade rate (and its potential connection with the direct vs inverse cascades) deserves to be adequately considered and is beyond the scope of the present paper. Finally, for the statistical convergence of the energy cascade rate, we analysed in detailed the correlation functions to assure its convergence, i.e. the constant value in the MHD scales). Further discussion is described in the Appendix \ref{appe}. 

%\ADD{Moreover, in order to assure the statistical convergence of the third-order moment, we used one day of events of different time duration (i.e., different window width) between 30 minutes and 12 hours so as to measure the fluctuations of the incompressible energy cascade rate over the MHD scales (not shown in this paper). Indeed, we observed that the energy cascade approaches a constant value in the MHD scale for 1 hour event. Besides, we found a similar behavior taking the average of the cascade rate over each window. In particular, this value is the average of the mean of the energy cascade over every window studied. There are several tests of convergence of the incompressible \citep[e.g.,][]{Wang2022,Co2015} and compressible \citep{H2017a} cascade rate that can be addressed, although this is beyond the scope of this paper. CHECKEAR ESTO}  

\subsection{The compressible and incompressible energy cascade rates}

\begin{figure}[ht]
    \centering
    \includegraphics[width=0.85\linewidth]{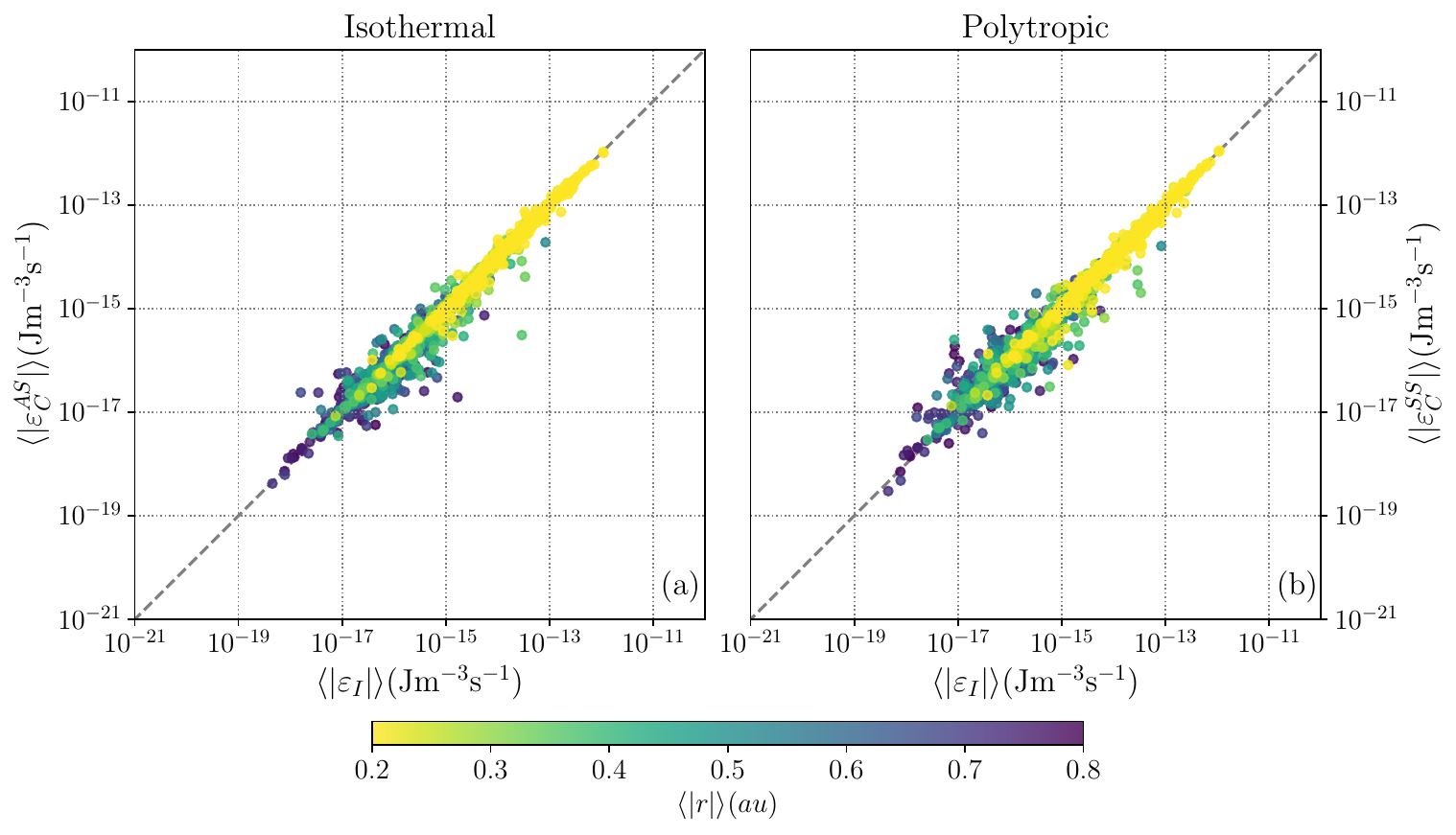}
    \caption{Total compressible energy cascade rate (absolute values) in the MHD scales as a function of the incompressible ones in the case of using (a) an isothermal (AS) and (b) polytropic (SS) model, respectively. The colobar corresponds to the heliocentric distance per event.}
    \label{R_total_INC_grid}
\end{figure}

\begin{figure}[ht]
    \centering
    \includegraphics[width=0.85\linewidth]{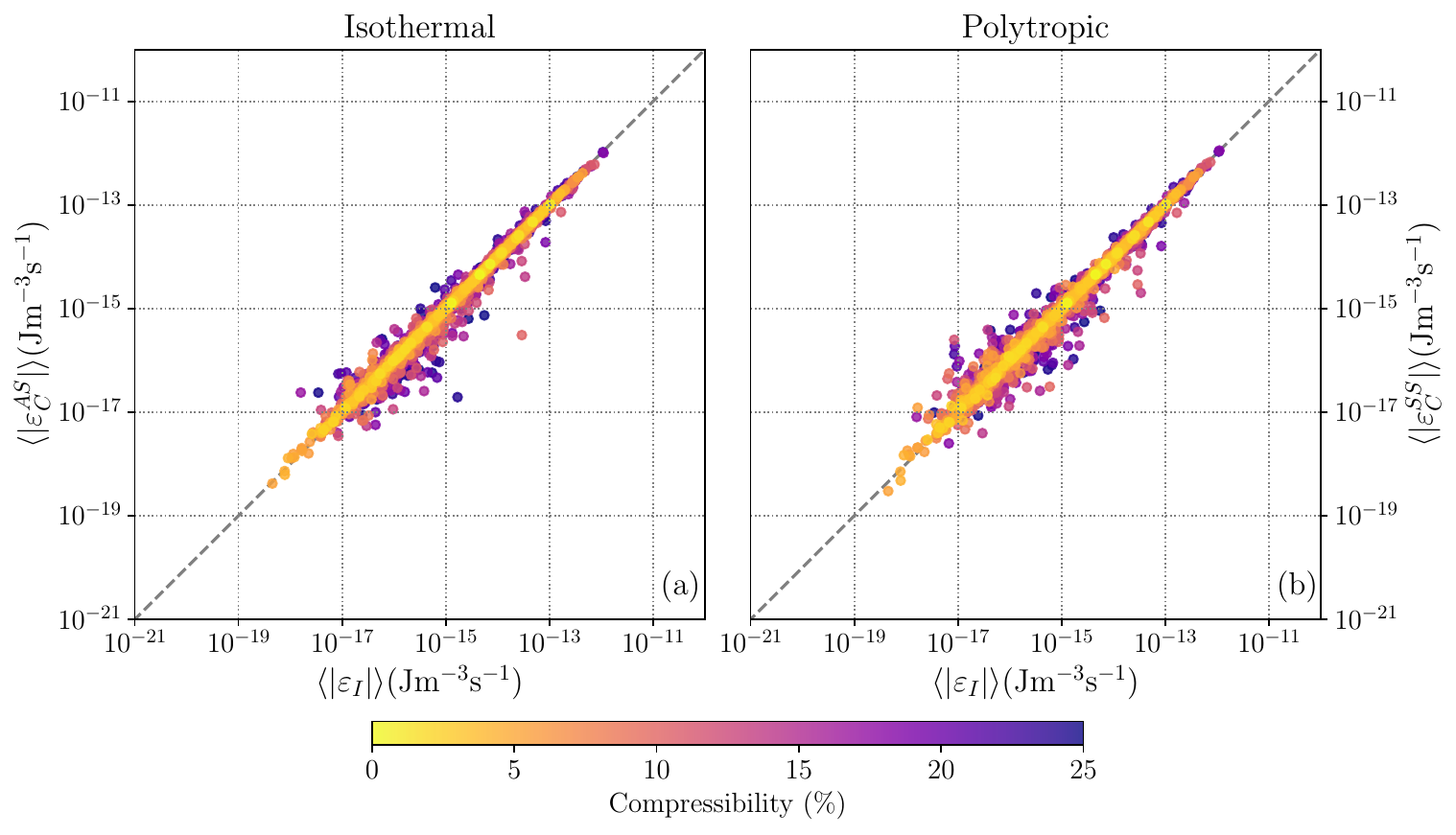}
    \caption{Total compressible energy cascade rate (absolute values) in the MHD scales as a function of the incompressible ones in the case of using (a) an isothermal (AS) and (b) polytropic (SS) model, respectively. The colobar corresponds to the level of compressibility (i.e., $\sqrt{\langle\rho^2\rangle-\langle\rho\rangle^2}/\langle\rho\rangle$) per event.}
    \label{Compresibilidad_total_INC}
\end{figure}

Figure \ref{R_total_INC_grid} shows the absolute value of the total compressible energy cascade rate $\langle|\varepsilon_C|\rangle$ as a function of the absolute value of the incompressible cascade rate $\langle|\varepsilon_I|\rangle$ for (a) the isothermal (superscript {\it AS} from \citet{A2017b}) and (b) the polytropic (superscript {\it SS} from \citet{S2021}) model, respectively, using the heliocentric distance per event as the colorbar. We found a strong correlation between the cascade rate amplitude and the distance to the Sun. In particular, the closer to the Sun PSP is, the larger the energy cascade amplitude is. This is in agreement with previous work in the solar wind studying the incompressible energy cascade rate \citep[e.g.,][]{B2020,A2022}. Likewise, Figure \ref{Compresibilidad_total_INC} shows the absolute value of the total compressible vs the incompressible energy cascade rate yet using the compressibility rate $\sqrt{\langle\rho^2\rangle-\langle\rho\rangle^2}/\langle\rho\rangle$ (in percent) of each event as the colorbar. We observed an increase in the energy cascade rate as the level of compressibility grows, reaching up to 25$\%$. Also, we note a clear correlation between the compressible and the incompressible cascade rates. In fact, given the weakly compressible nature of the solar wind, it is not surprising that these two energy cascades are similar and close to equality. Moreover, we observed (see Section \ref{B}) that the first term of equation \eqref{Compre}, i.e. the Yaglom-like term, dominates over the second one. As we mentioned above, this term tends to the incompressible expression in the limit of constant density. However, it is worth mentioning that, for the lowest values of the cascade rates in our data set (at $\sim$ 0.8 au), there is a slight increase of the compressible cascade with respect to the incompressible one. It is worth noting that, as we are studying the total expression \eqref{Compre}, we are not able to differentiate between the first and the second term which are the Yaglom-like term and the purely compressible one, respectively. 
%As a result, as we will see from the next section, the dominant term in the energy cascade is the Yaglom-like term 
%

\subsection{The isothermal and polytropic energy cascade components}\label{B}

\begin{figure}[ht]
    \centering
    \includegraphics[width=0.85\linewidth]{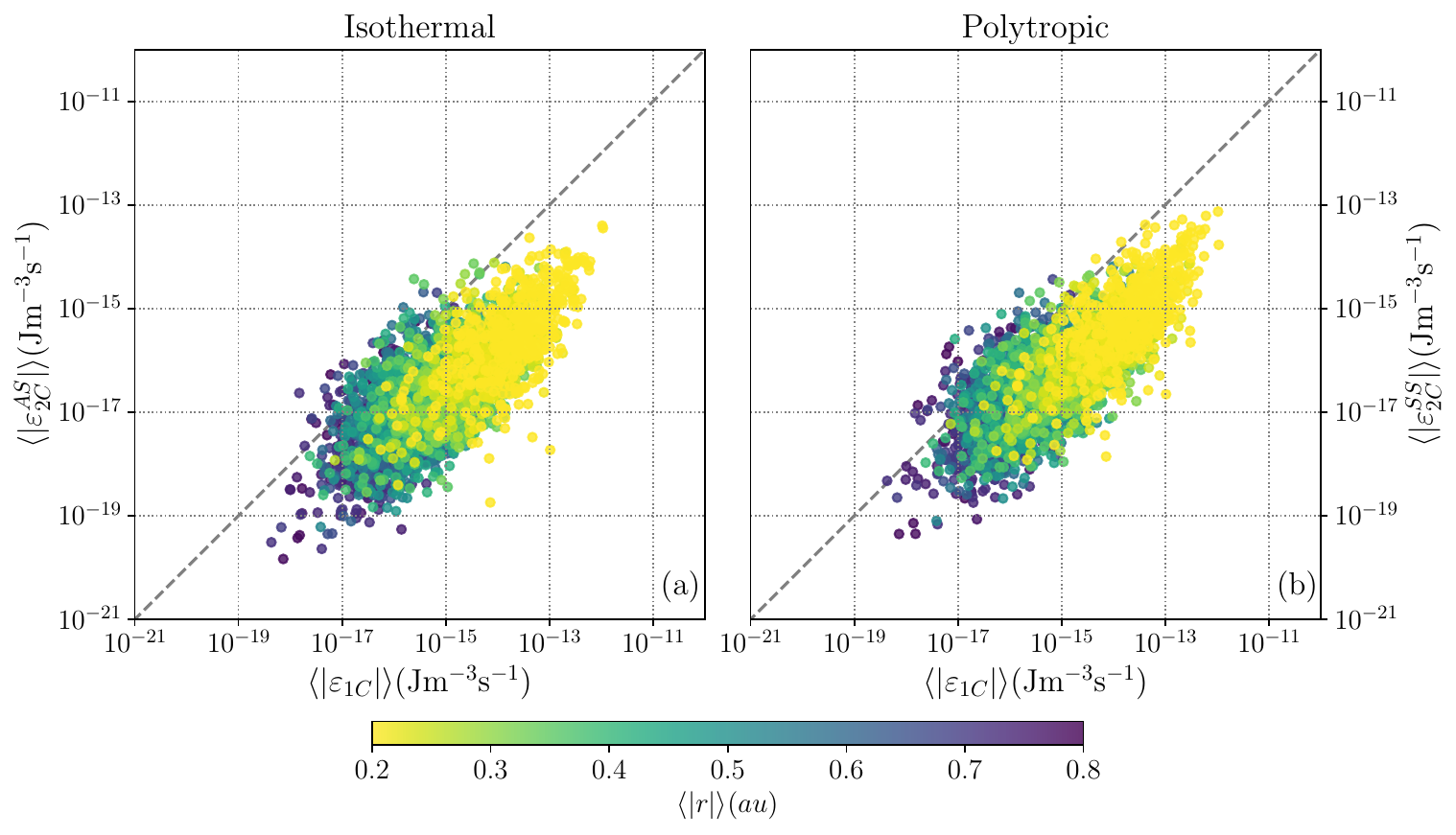}
    \caption{The compressible cascade rate component $\langle|\varepsilon_{2C}|\rangle$ as a function of the Yaglom-like term $\langle|\varepsilon_{1C}|\rangle$ in the MHD scales in the case of using (a) an isothermal (AS) and (b) polytropic (SS) model, respectively. The colobar corresponds to the Heliocentric distance per event.}
    \label{R_2_1_inc_grid}
\end{figure}

\begin{figure}[ht]
    \centering
    \includegraphics[width=0.85\linewidth]{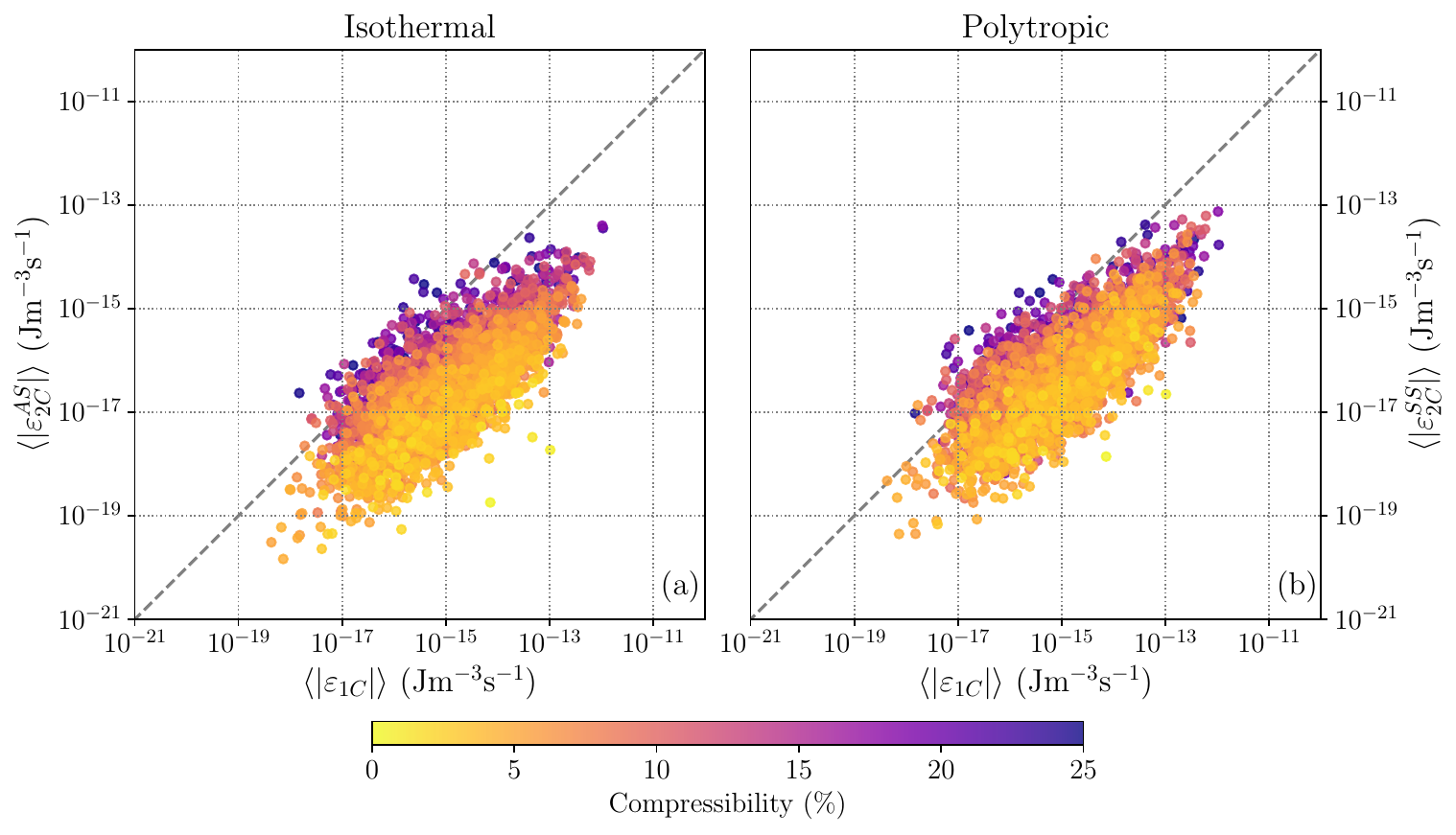}
    \caption{The compressible cascade rate component $\langle|\varepsilon_{2C}|\rangle$ as a function of the Yaglom-like term $\langle|\varepsilon_{1C}|\rangle$ in the MHD scales in the case of using (a) an isothermal (AS) and (b) polytropic model (SS), respectively. The colobar corresponds to the level of compressibility per event.}
    \label{Compresibilidad_2_1_inc}
\end{figure}

Figure \ref{R_2_1_inc_grid} and \ref{Compresibilidad_2_1_inc} show the absolute value of (a) the isothermal component $\langle|\varepsilon_{2C}^{AS}|\rangle$ and (b) the polytropic component $\langle|\varepsilon_{2C}^{SS}|\rangle$ (see Eq.~\ref{com_flux}) as a function of the Yaglom compressible generalization component $\langle|\varepsilon_{1C}|\rangle$ (see Eq.~\ref{yaglom}), respectively. Again, the colorbar corresponds to the heliocentric distance and the compressibility percent, respectively. Although, in general, the dominant term in the MHD scales is $\langle|\varepsilon_{1C}|\rangle$, we noticed that even for the nearly incompressible solar wind (Compressibility $\le\%10$), for some events the purely compressible term is relevant for an adequate estimation of the total cascade. We also observed that there are some events where the energy of the compressible term exceeds the Yaglom-like term. Moreover, the level of compressibility increases up to 20-25$\%$ in those events. Therefore, when we compared the compressible and the incompressible energy cascade in Figures \ref{Compresibilidad_total_INC} and \ref{Compresibilidad_2_1_inc}, there is an impact of density fluctuation in the compressible components, even though most of the events are in a quasi incompressible state.

In the isothermal and polytropic components, we found the same behavior of the total compressible energy cascade with the heliocentric distance (as seen in Figure \ref{R_total_INC_grid}). Thus, we observe the fact that the energy cascade rates, both compressible and incompressible ones, are ordered by the heliocentric distance. Regarding the polytropic and the isothermal models, we noted that there is also a slight increase of the compressible term in the first case with respect to the second one. In addition, we note that there are larger cascade values when we use the polytropic model.

\subsection{The effect of compressibility and heliocentric distance over the energy cascade rates}

\begin{figure}[]
    \centering
    {\includegraphics[width=0.49\linewidth]{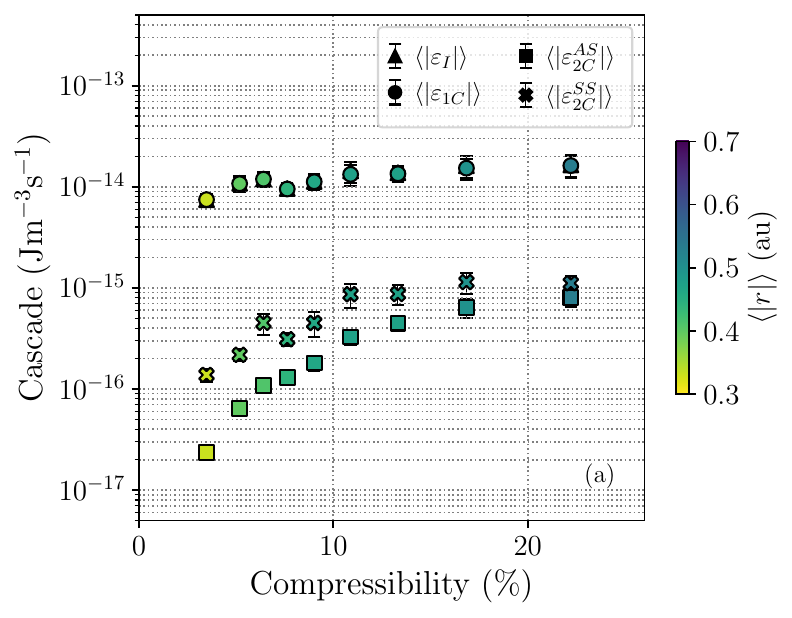}}{\includegraphics[width=0.49\linewidth]{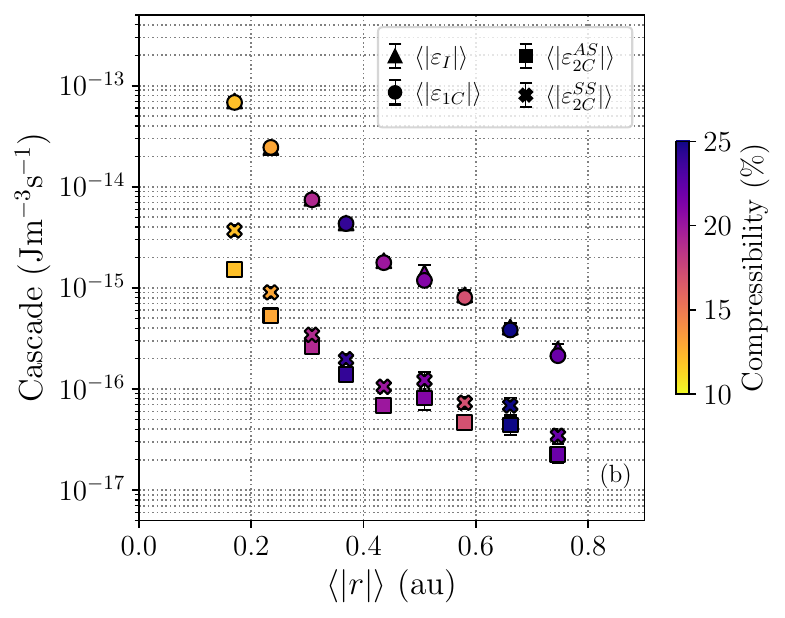}}
    \caption{Average of the energy cascade rate for a given: (a) compressibility bin using a colorbar that corresponds to the heliocentric distance and (b) a heliocentric distance bin where the colorbar corresponds to the level of compressibility. The compressible isothermal and polytropic components ($\langle|\varepsilon_{2C}^{AS}|\rangle$ and $\langle|\varepsilon_{2C}^{SS}|\rangle$, resp.), the Yaglom generalization term ($\langle|\varepsilon_{1C}|\rangle$) and the incompressible cascade ($\langle|\varepsilon_{I}|\rangle$) are included.}
    \label{bineados}
\end{figure}

Figure \ref{bineados} shows the bin-average absolute value of the energy cascade rate components as a function of (a) the compressibility and (b) the heliocentric distance. It is worth mentioning that we group events according to these two magnitudes and then, we segment and sort data values into bins. Finally, we took the average value of each bin. Moreover, in Fig.~\ref{bineados} (a) the colorbar corresponds to $\langle|r|\rangle$ and in Fig.~\ref{bineados} (b) the colorbar corresponds to the compressibility. In particular, we compared the behavior of the total expression of the compressible cascade, the compressible components, and the incompressible cascade. We reported the standard error as the error bars. Figure \ref{bineados} summarizes the results seen in previous Figs.~\ref{R_total_INC_grid}-\ref{Compresibilidad_2_1_inc}. 

As we discussed previously (in Figure \ref{R_2_1_inc_grid}), we notice that as we approach to the Sun, the absolute value of the compressible energy cascade rate increases. Also, as the level of compressibility increases in the solar wind (see Figure \ref{Compresibilidad_2_1_inc}), there is an important increase of the \ADD{absolute value of the} compressible energy cascade, approximately one order of magnitude. In the case of the incompressible cascade rate, the increment of the rate is less relevant compared to the compressible one. It is worth mentioning that, there is a clear relation between the compressibility and the distance to the Sun: as we travel away from the Sun, the larger the level of compressibility is. Interestingly, here we show that, on average, $\langle|\varepsilon_{2C}^{SS}|\rangle > \langle|\varepsilon_{2C}^{AS}|\rangle$ since $76\%$ of the events satisfy this condition. %Meanwhile, only $23\%$ of the events have $\langle|\varepsilon_{2C}^{SS}|\rangle < \langle|\varepsilon_{2C}^{SS}|\rangle$.

\subsection{\ADD{The sign of the incompressible and compressible energy cascade rates}}\label{Sign}

\begin{figure}[ht]
    \centering
    \includegraphics[width=0.85\linewidth]{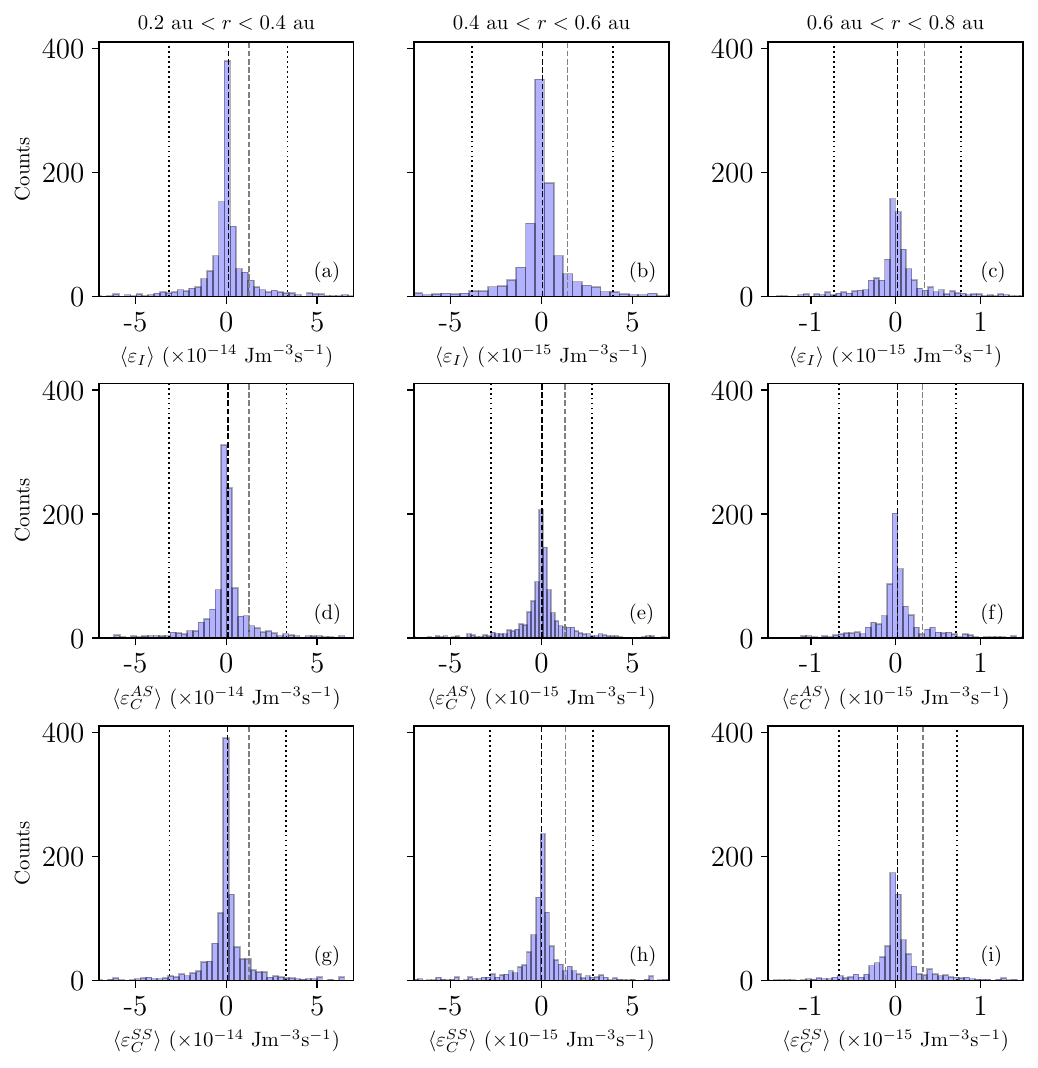}
    \caption{\ADD{Histograms of the total signed (a-c) $\langle\varepsilon_I\rangle$, (d-f) $\langle\varepsilon^{AS}_C\rangle$ and (g-i) $\langle\varepsilon^{SS}_C\rangle$ distribution as a function of different heliocentric distances. Black dashed and dotted lines correspond to the mean and the standard deviation of the signed cascade, respectively. For the sake of comparison we have added in gray dashed line the mean of the absolute value of the energy cascade rate.}}
    \label{Histograma_signo}
\end{figure}

\begin{figure}[ht]
    \centering
    \includegraphics[width=0.85\linewidth]{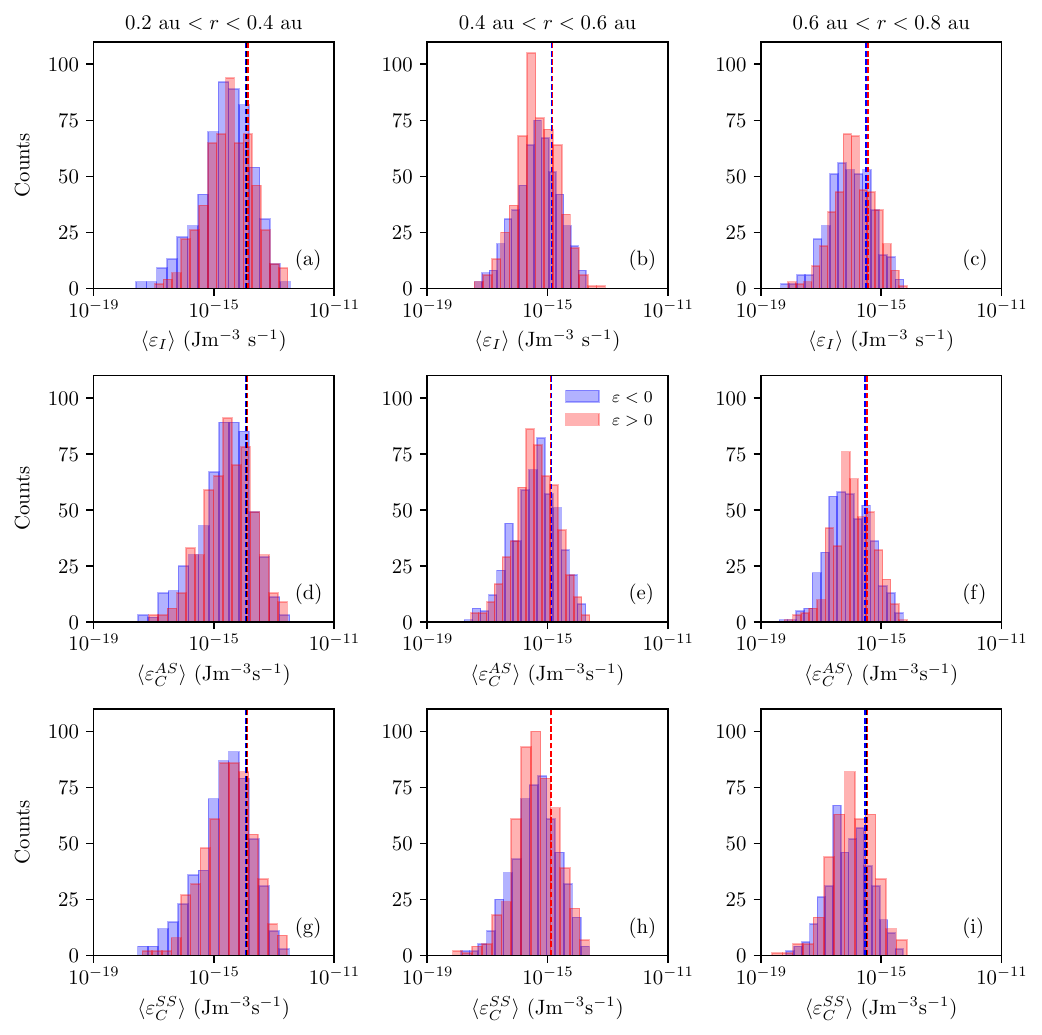}
    \caption{\ADD{Histograms of the signed (a-c) $\langle\varepsilon_I\rangle$, (d-f) $\langle\varepsilon^{AS}_C\rangle$ and (g-i) $\langle\varepsilon^{SS}_C\rangle$ distribution as a function of different heliocentric distances. Red and blue bars correspond to positive and negative values, respectively. Finally, red, blue and black dash lines correspond to the mean of the positive, negative and absolute values, respectively.}}
    \label{Histograma_signo_pos_neg}
\end{figure}

\ADD{In the previous sections, we have investigated the absolute value of the incompressible and compressible energy cascade rate. To investigate the connection between the absolute value of the cascade and the real cascade rate, we study the signed energy cascade rate in this subsection Figure \ref{Histograma_signo} and \ref{Histograma_signo_pos_neg} show the signed (a-c) incompressible, (d-f) isothermal and (g-i) polytropic energy cascade rate distributions for different heliocentric distances $r$. On the one hand, in Figure \ref{Histograma_signo}, we compared the signed and the absolute value of the energy cascade. To do this, we have added in black dashed and dotted lines the mean value and the standard deviation of the signed energy cascade, respectively. Besides, we have included in a gray dashed line the mean of the absolute value of each distribution. In general, we observed that the mean of the absolute value of the cascade exceed the mean of the signed cascade. Therefore, it is worth mentioning that there is a limitation in the analysis of the absolute value of $\varepsilon$ since it is artificially increasing due to the use of the absolute value.}

\ADD{In Figure \ref{Histograma_signo_pos_neg}, we studied separately the amount of positive (red bars) and negative (blue bars) values of the energy cascade. For the sake of comparison, we have moved the negative values distribution to the positive axis using the absolute value of them. For each particular distribution, we added in red, blue and black dash lines the mean of the positive, negative and absolute values, respectively. We observed similar means for the positive and negative values for each distribution (similar results are obtained using the mean value of each distribution). Moreover, we found that the incompressible and compressible energy cascade rates have statistically the same amount of negative and positive values in our data set. On the other hand, the sign of the cascade is related to the direct and inverse cascade through the Taylor hypothesis \ADD{\citep[e.g.,][]{So2007, Ma2023}}. In particular, a positive (respectively negative) value of $\varepsilon$ imply a direct (inverse) energy cascade. Therefore, in Figure \ref{Histograma_signo_pos_neg} we obtained a positive sign of the mean value of the incompressible and compressible (isothermal and polytropic) cascade which indicates a direct cascade at different heliocentric distances, despite having negative values of the energy cascade rate for individual hourly events.}

\begin{figure}[ht]
    \centering
    \includegraphics[width=0.85\linewidth]{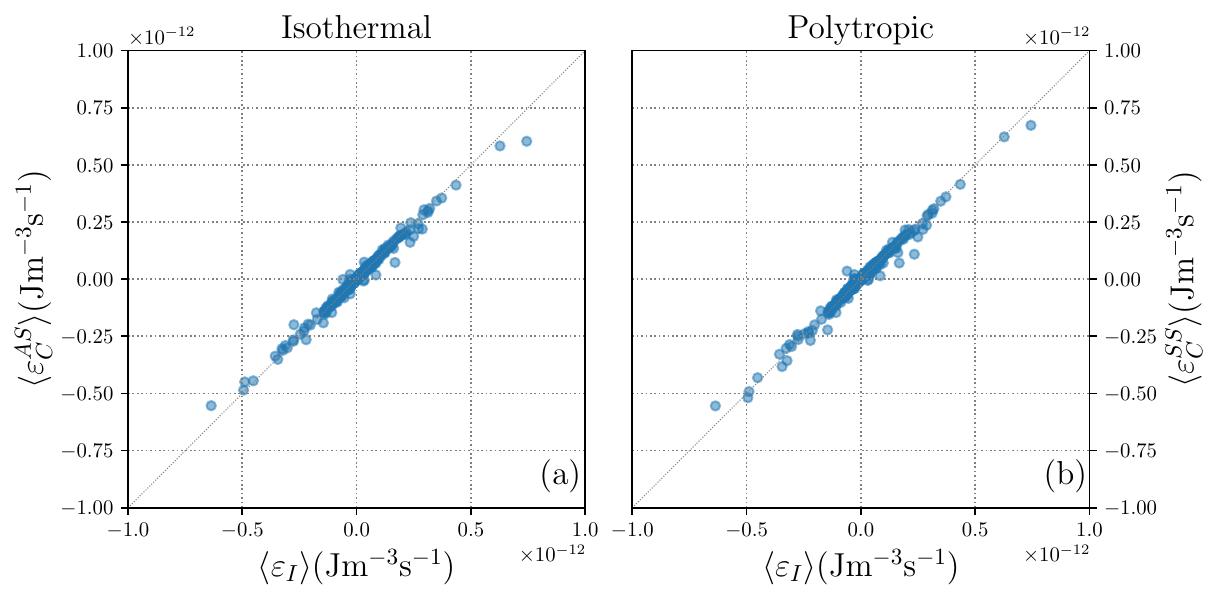}
    \caption{\ADD{The signed compressible energy cascade rates (a) isothermal and (b) polytropic as a function of the incompressible energy cascade rates.}}
    \label{cascada_signo}
\end{figure}

\ADD{Finally, we studied the relative sign between the compressible and incompressible cascade rates.} Figure \ref{cascada_signo} shows the signed compressible cascade rate as a function of the signed incompressible one. We observed the same sign for both cascades in most of the events ($\sim 97\%$) for the averaged incompressible and compressible energy cascade rates. We concluded that for both isothermal and polytropic models, the compressible fluctuations do not affect the particular direction of the cascade rate.

%The presence of positive and negative values for the cascade rate have been observed in solar wind turbulence \citep{Co2015, St2011,St2010}. Using Advanced Composition Explorer (ACE) observations, \citet{Co2015} have found evidence for intermittent dynamics with a high level of variability relative to the mean cascade rates. In particular, the average cascade rate reported at 1 au is compatible to our results for large heliocentric distances

%This is in good agreement with the previous observational results reported by \cite{H2017a}. In particular, the authors showed that the correlation between the signed incompressible and reduced isothermal compressible cascades have the same sign for most of their cases.


\section{Discussion and Conclusions}\label{sec:Discussion}

In the present study, we estimated the \ADD{absolute value of the} energy cascade rate from both incompressible and compressible cases, using two different closures for the compressible model, i.e., the isothermal and the polytropic case. In particular, we computed the \ADD{absolute value of the} incompressible and compressible energy cascade rate in the solar wind at different heliocentric distances ($\sim 0.1 -  0.8$ au). Firstly, we found a clear correlation between the energy cascade rate and the distance to the Sun (Figures \ref{R_total_INC_grid} and \ref{R_2_1_inc_grid}). Our observational results show that, as we get closer to the Sun, the absolute value of the energy cascade rate increase. This observational result is compatible with recent studies \citep{Ad2015,B2020,Ad2021,A2021,A2022}. For instance, \citet{A2022} showed a correlation between the incompressible energy cascade and the heliocentric distance in the inner heliosphere using isotropic and anisotropic exact relations. Also, \citet{B2020} have estimated the energy transfer rate from the first PSP perihelion using an incompressible exact relation and the von-Kármán decay law. The authors found that the energy cascade obtained near the perihelion is about 100 times higher than the average value at 1 au. Although we used a more complex theoretical model, as we include density fluctuations to model the expression for the energy cascade, our results show a similar increase of the \ADD{absolute value of the} incompresssible and compressible energy cascade rate (up to two order of magnitude) as we approach to the Sun, probably due to the increase of the mean value of the amplitude of the fluctuations of plasma density, magnetic and velocity field.

Secondly, we observed an increase of the \ADD{absolute value of the} compressible energy cascade rate when the level of compressibility increases in the solar wind. We analyzed the competition between the Yaglom-like term (i.e., Eq.~\eqref{yaglom}) and the purely compressible term (i.e., Eq.~\eqref{com_flux}) in the total expression of the compressible cascade for both isothermal and polytropic models (see Figures \ref{Compresibilidad_2_1_inc} and \ref{R_2_1_inc_grid}). Despite the fact that the first term is dominant in most of the cases, we found that the second term (purely compressible) still may plays a relevant role for a proper estimation of the total compressible cascade in more compressible environments. Moreover, we obtained that, for the most compressible events ($20\%$-$25\%$), there is  a slightly growth in the compressible cascade with respect to the incompressible one. Therefore, higher density fluctuations in the plasma lead to increasing $\varepsilon_C$ over $\varepsilon_I$ \citep{So2007,M2008,A2019b,A2021}. Similarly, using THEMIS, MAVEN and PSP observations (at the first encounter), \citet{A2021} reported moderate increases of the isothermal compressible cascade with respect to the incompressible one at different heliocentric distances. In our case of study, we expand these previous results including the polytropic model and much more extended data set. Our observational results showed more significant increments between the compressible and incompressible cascade, especially for the events with smaller cascade energy values. In addition, \citet{H2017a} showed that the energy cascade rate increases as compressibility increases in the plasma in the slow solar wind. Note that in the present work, almost all the events are in the slow solar wind ($94 \%$) so we confirmed this previous observational results.

Then, we reported the average of the \ADD{absolute value of the} energy cascade rate in order to study and compare the behavior between the compressibility levels and the heliocentric distance (in Figure \ref{bineados}). We related these two magnitudes and found out that the compressibility increases as we increase the distance to the Sun. Our results are in good agreement with \citet{Ch2020}, who computed the magnetic compressibility coefficient $C_B = (\delta|B|/|\delta B|)^2$ and showed a clear decrease toward smaller heliocentric distances. They observed that the compressibility levels at PSP perihelion are an order of magnitude smaller at 1 au. However, their results are based on magnetic compressibility in the frequency domain unlike the present work where we computed the mean values in the real space. On the other hand, \citet{Ad2020} studied the frequency distribution of the solar wind compressibility between $0.17$ au and $0.61$ au and showed that density fluctuation distribution is concentrated mainly around $\sim$ 0.15 au and decreases with the heliocentric distance. This result is compatible with the density fluctuation levels found by \citet{A2021}. Instead, we showed that the compressibility distribution reach its highest values approximately at 0.5 au and decreases as we get closer to the Sun. 

\ADD{In addition, we calculated the absolute value of the} compressible energy cascade by using an isothermal and polytropic model based on previous theoretical works (\citet{A2017b,S2021}). From Figures \ref{R_2_1_inc_grid}, \ref{Compresibilidad_2_1_inc} and \ref{bineados} mentioned above and, specially, Figure \ref{bineados}, we noticed that there is a clear trend of the polytropic cascade to be larger than the isothermal one. In particular, \citet{S2021} observed the same behavior at MHD scales despite using few events unlike our present statistical work. However, it is worth mentioning that the two compressible models (isothermal and polytropic) give essentially the same cascade rate due to the fact that the contribution of the Yaglom-like term $\varepsilon_{1C}$ tends to be dominate over $\varepsilon_{2C}$ in the inertial MHD range in both cases. 

\ADD{We carried out an study of the sign of the energy cascade rate by analyzing the signed incompressible and compressible cascade distributions for different heliocentric distances. First, we compared the signed cascade with its absolute value distributions and as we expect there is a limitation in the analysis of the absolute value of the cascade rate since the absolute value over estimates the cascade values and these values could not be representative of the real cascade rate. Second, our observational results show statistically the same amount of positive and negative values in our data set. In fact, the mean value of these distributions are quite similar. Moreover, in all the distributions we found a positive sign for the mean value of signed energy cascade, which indicates a connection with a direct turbulent cascade process. The presence of positive and negative values of the cascade rate have been investigated previously in solar wind turbulence \citep{Co2015, St2011,St2010}. Using Advanced Composition Explorer (ACE) observations, \citet{Co2015} have found evidence for intermittent dynamics with a high level of variability relative to the mean cascade rates. In particular, the average cascade rate reported at 1 au is compatible to our results for large heliocentric distances. Finally, we included a brief study of the relative sign of the compressible and incompressible cascade rate values and we found that, in general, the compressible fluctuations do not affect the direction of the cascade rate. This is in good agreement with the previous observational results reported by \citet{H2017a}. In particular, the authors showed that the correlation between the signed incompressible and reduced isothermal compressible cascades have the same sign for most of their cases at 1 au.}

In summary, our observational results support the idea that the \ADD{absolute value of the} compressible energy cascade rate increases at small heliocentric distance and at large values of compressibility or density fluctuation levels. Moreover, \ADD{the absolute value of} the compressible energy cascade rate increases at small distances, while the fraction of the compressible cascade rate to the incompressible cascade rate seems to decrease as we move away from the Sun. This is a direct consequence from the fact that density fluctuations (or compressibility) decrease as we increase the heliocentric distance. Nevertheless, there are some aspects of this work that require improvement. First, we found some discrepancies about the intrinsic relation between the heliocentric distances and the compressibility levels. Previous observational results \citep{A2021} have found that at smaller heliocentric distances, the compressibility and density fluctuations tends to be larger. Therefore, some other process may be acting to suppress/increase fluctuations as we get closer to the Sun. Second, in the present work we only focus in the MHD scales. Recent exact relations could be used to estimate the transfer of energy as we reach the sub ion scales \citep{A2018,S2021,S2022}. In forthcoming works, further statistical investigation of these topics are going to be carried out using in situ data in more compressible environments (like the Earth's magnetosheath). 

\begin{acknowledgements}
N.A.~and P.D.~acknowledge financial support from CNRS/CONICET Laboratoire International Associé (LIA) MAGNETO. N.A.~acknowledges financial support from the following grants: PICT 2018-1095 and UBACyT 20020190200035BA. P.D.~acknowledges financial support from PIP Grant No.~11220150100324
and 11220200101752 and PICT Grant No.~2018-4298. The data that support the findings of this study are openly available in Parker Solar Probe Data Repository [\cite{1}].)
\end{acknowledgements}

\section{Appendix: The convergence of the energy cascade rate}\label{appe}
\begin{figure}[ht]
    \centering
    \includegraphics[width=0.85\linewidth]{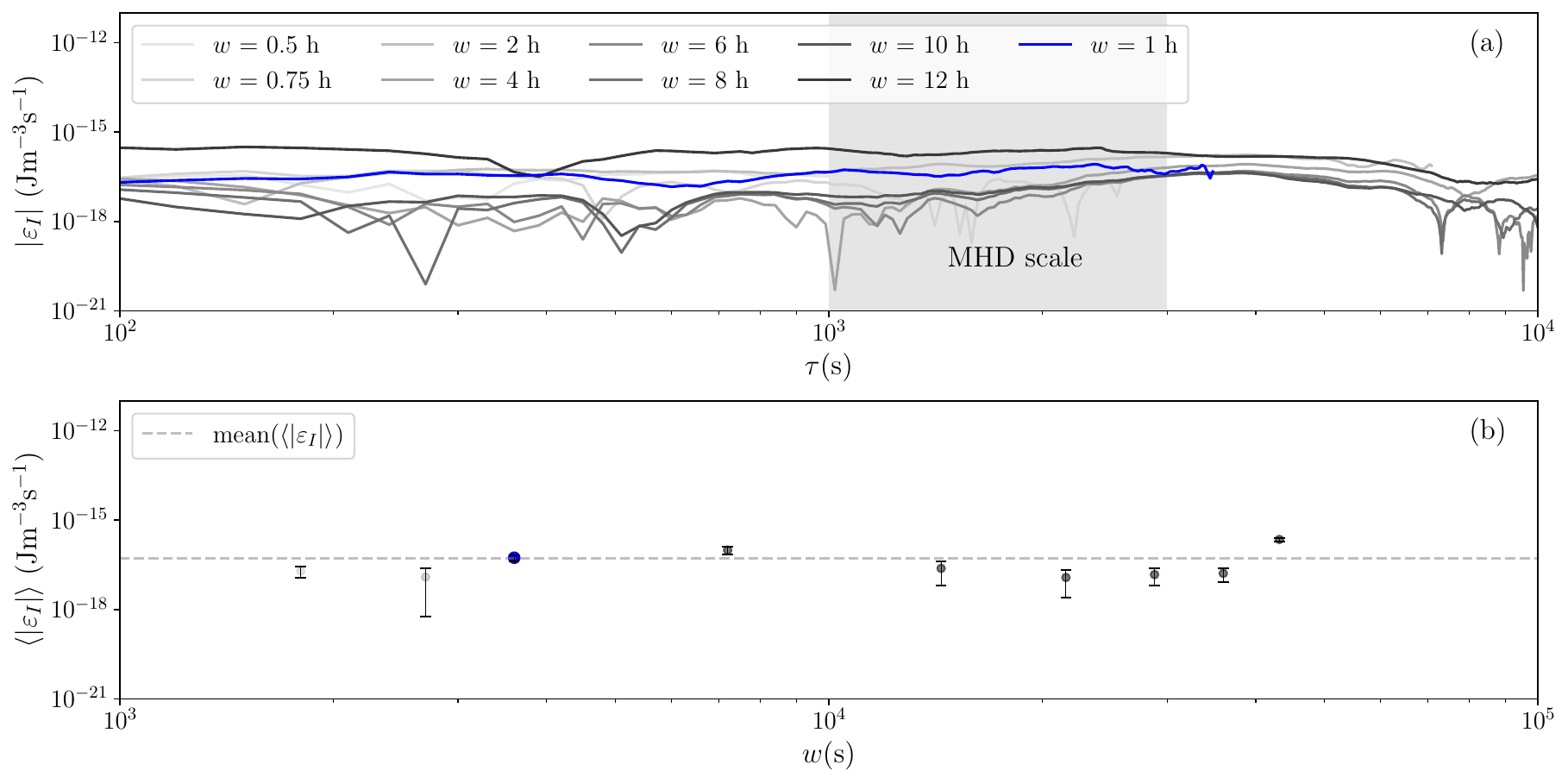}
    \caption{(a) Absolute value of the incompressible energy cascade rate for one day of events of different time duration (width). (b) Average of the energy cascade for each event.}
    \label{Convergence}
\end{figure}

We studied the statistical convergence of the third-order moment technique and for some randomly chosen events, we used different time duration lengths to measure how constant the absolute value of the incompressible energy cascade rate actually is in the MHD scales. Figure \ref{Convergence} shows the average of the energy cascade rate in the MHD scales for each particular window (between 30 minutes and 12 hours) and we observed that, indeed, the cascade rate approaches a constant value compatible with the average at 1 hour.

In fact, the convergence of the energy cascade rate using third-order theory becomes relevant to provide a local measurement of the nonlinear terms. Finally, different tests of convergence of the incompressible \citep[e.g.,][]{Wang2022,Co2015,St2009} and compressible \citep{H2017a} energy cascade rate have been applied to investigate the convergence of this technique.

\bibliography{cites.bib}

\end{document}

%% file: inputs_mark.tex
\usepackage{graphicx}
\usepackage{txfonts}
\usepackage{natbib}
\usepackage{subfigure}
\usepackage{graphicx, color}
\usepackage[english]{babel}
\draft % marks overfull lines with a black rule on the right
\def\ADD#1{{{#1}}}  
\usepackage[hidelinks,colorlinks=true,linkcolor=blue,citecolor=blue,urlcolor=blue]{hyperref}
\usepackage{lineno}

%% file: main_mark.bbl
%merlin.mbs aipauth4-1.bst 2010-07-25 4.21a (PWD, AO, DPC) hacked
%Control: author (9) reversed initials
\begin{thebibliography}{82}%
\makeatletter
\providecommand \@ifxundefined [1]{%
 \@ifx{#1\undefined}
}%
\providecommand \@ifnum [1]{%
 \ifnum #1\expandafter \@firstoftwo
 \else \expandafter \@secondoftwo
 \fi
}%
\providecommand \@ifx [1]{%
 \ifx #1\expandafter \@firstoftwo
 \else \expandafter \@secondoftwo
 \fi
}%
\providecommand \natexlab [1]{#1}%
\providecommand \enquote  [1]{``#1''}%
\providecommand \bibnamefont  [1]{#1}%
\providecommand \bibfnamefont [1]{#1}%
\providecommand \citenamefont [1]{#1}%
\providecommand \href@noop [0]{\@secondoftwo}%
\providecommand \href [0]{\begingroup \@sanitize@url \@href}%
\providecommand \@href[1]{\@@startlink{#1}\@@href}%
\providecommand \@@href[1]{\endgroup#1\@@endlink}%
\providecommand \@sanitize@url [0]{\catcode `\\12\catcode `\$12\catcode
  `\&12\catcode `\#12\catcode `\^12\catcode `\_12\catcode `\%12\relax}%
\providecommand \@@startlink[1]{}%
\providecommand \@@endlink[0]{}%
\providecommand \url  [0]{\begingroup\@sanitize@url \@url }%
\providecommand \@url [1]{\endgroup\@href {#1}{\urlprefix }}%
\providecommand \urlprefix  [0]{URL }%
\providecommand \Eprint [0]{\href }%
\providecommand \doibase [0]{http://dx.doi.org/}%
\providecommand \selectlanguage [0]{\@gobble}%
\providecommand \bibinfo  [0]{\@secondoftwo}%
\providecommand \bibfield  [0]{\@secondoftwo}%
\providecommand \translation [1]{[#1]}%
\providecommand \BibitemOpen [0]{}%
\providecommand \bibitemStop [0]{}%
\providecommand \bibitemNoStop [0]{.\EOS\space}%
\providecommand \EOS [0]{\spacefactor3000\relax}%
\providecommand \BibitemShut  [1]{\csname bibitem#1\endcsname}%
\let\auto@bib@innerbib\@empty
%</preamble>
\bibitem [{1()}]{1}%
  \BibitemOpen
  \href@noop {} {\enquote {\bibinfo {title} {{NASA Parker Solar Probe}},}\
  }\bibinfo {howpublished}
  {\url{https://research.ssl.berkeley.edu/data/psp/}}\BibitemShut {NoStop}%
\bibitem [{\citenamefont {Adhikari}\ \emph {et~al.}(2015)\citenamefont
  {Adhikari}, \citenamefont {Zank}, \citenamefont {Bruno}, \citenamefont
  {Telloni}, \citenamefont {Hunana}, \citenamefont {Dosch}, \citenamefont
  {Marino},\ and\ \citenamefont {Hu}}]{Ad2015}%
  \BibitemOpen
  \bibfield  {author} {\bibinfo {author} {\bibnamefont {Adhikari},
  \bibfnamefont {L.}}, \bibinfo {author} {\bibnamefont {Zank}, \bibfnamefont
  {G.}}, \bibinfo {author} {\bibnamefont {Bruno}, \bibfnamefont {R.}}, \bibinfo
  {author} {\bibnamefont {Telloni}, \bibfnamefont {D.}}, \bibinfo {author}
  {\bibnamefont {Hunana}, \bibfnamefont {P.}}, \bibinfo {author} {\bibnamefont
  {Dosch}, \bibfnamefont {A.}}, \bibinfo {author} {\bibnamefont {Marino},
  \bibfnamefont {R.}}, \ and\ \bibinfo {author} {\bibnamefont {Hu},
  \bibfnamefont {Q.}},\ }\bibfield  {title} {\enquote {\bibinfo {title} {The
  transport of low-frequency turbulence in astrophysical flows. ii. solutions
  for the super-alfv\'enic solar wind},}\ }\href {\doibase
  10.1088/0004-637X/805/1/63} {\bibfield  {journal} {\bibinfo  {journal} {The
  Astrophysical Journal}\ }\textbf {\bibinfo {volume} {805:63}},\ \bibinfo
  {pages} {18pp} (\bibinfo {year} {2015})}\BibitemShut {NoStop}%
\bibitem [{\citenamefont {Adhikari}\ \emph {et~al.}(2021)\citenamefont
  {Adhikari}, \citenamefont {Zank}, \citenamefont {Zhao}, \citenamefont
  {Telloni}, \citenamefont {Horbury}, \citenamefont {O'Brien}, \citenamefont
  {Evans}, \citenamefont {Angelini}, \citenamefont {Owen}, \citenamefont
  {Louarn},\ and\ \citenamefont {Fedorov}}]{Ad2021}%
  \BibitemOpen
  \bibfield  {author} {\bibinfo {author} {\bibnamefont {Adhikari},
  \bibfnamefont {L.}}, \bibinfo {author} {\bibnamefont {Zank}, \bibfnamefont
  {G.}}, \bibinfo {author} {\bibnamefont {Zhao}, \bibfnamefont {L.}}, \bibinfo
  {author} {\bibnamefont {Telloni}, \bibfnamefont {D.}}, \bibinfo {author}
  {\bibnamefont {Horbury}, \bibfnamefont {T.}}, \bibinfo {author} {\bibnamefont
  {O'Brien}, \bibfnamefont {H.}}, \bibinfo {author} {\bibnamefont {Evans},
  \bibfnamefont {V.}}, \bibinfo {author} {\bibnamefont {Angelini},
  \bibfnamefont {V.}}, \bibinfo {author} {\bibnamefont {Owen}, \bibfnamefont
  {C.}}, \bibinfo {author} {\bibnamefont {Louarn}, \bibfnamefont {P.}}, \ and\
  \bibinfo {author} {\bibnamefont {Fedorov}, \bibfnamefont {A.}},\ }\bibfield
  {title} {\enquote {\bibinfo {title} {Evolution of anisotropic turbulence in
  the fast and slow solar wind: Theory and solar orbiter measurements},}\
  }\href@noop {} {\bibfield  {journal} {\bibinfo  {journal} {Astronomy and
  Astrophysics}\ } (\bibinfo {year} {2021})}\BibitemShut {NoStop}%
\bibitem [{\citenamefont {Adhikari}\ \emph {et~al.}(2020)\citenamefont
  {Adhikari}, \citenamefont {Zank}, \citenamefont {Zhao}, \citenamefont
  {Kasper}, \citenamefont {Korreck}, \citenamefont {Stevens}, \citenamefont
  {Case}, \citenamefont {Whittlesey}, \citenamefont {Larson}, \citenamefont
  {Livi} \emph {et~al.}}]{Ad2020}%
  \BibitemOpen
  \bibfield  {author} {\bibinfo {author} {\bibnamefont {Adhikari},
  \bibfnamefont {L.}}, \bibinfo {author} {\bibnamefont {Zank}, \bibfnamefont
  {G.~P.}}, \bibinfo {author} {\bibnamefont {Zhao}, \bibfnamefont {L.-L.}},
  \bibinfo {author} {\bibnamefont {Kasper}, \bibfnamefont {J.~C.}}, \bibinfo
  {author} {\bibnamefont {Korreck}, \bibfnamefont {K.~E.}}, \bibinfo {author}
  {\bibnamefont {Stevens}, \bibfnamefont {M.}}, \bibinfo {author} {\bibnamefont
  {Case}, \bibfnamefont {A.~W.}}, \bibinfo {author} {\bibnamefont {Whittlesey},
  \bibfnamefont {P.}}, \bibinfo {author} {\bibnamefont {Larson}, \bibfnamefont
  {D.}}, \bibinfo {author} {\bibnamefont {Livi}, \bibfnamefont {R.}},  \emph
  {et~al.},\ }\bibfield  {title} {\enquote {\bibinfo {title} {Turbulence
  transport modeling and first orbit parker solar probe (psp) observations},}\
  }\href@noop {} {\bibfield  {journal} {\bibinfo  {journal} {The Astrophysical
  Journal Supplement Series}\ }\textbf {\bibinfo {volume} {246}},\ \bibinfo
  {pages} {38} (\bibinfo {year} {2020})}\BibitemShut {NoStop}%
\bibitem [{\citenamefont {Alexandrova}\ \emph {et~al.}(2013)\citenamefont
  {Alexandrova}, \citenamefont {Chen}, \citenamefont {Sorriso-Valvo},
  \citenamefont {Horbury},\ and\ \citenamefont {Bale}}]{A2013}%
  \BibitemOpen
  \bibfield  {author} {\bibinfo {author} {\bibnamefont {Alexandrova},
  \bibfnamefont {O.}}, \bibinfo {author} {\bibnamefont {Chen}, \bibfnamefont
  {C.~H.~K.}}, \bibinfo {author} {\bibnamefont {Sorriso-Valvo}, \bibfnamefont
  {L.}}, \bibinfo {author} {\bibnamefont {Horbury}, \bibfnamefont {T.~S.}}, \
  and\ \bibinfo {author} {\bibnamefont {Bale}, \bibfnamefont {S.~D.}},\
  }\bibfield  {title} {\enquote {\bibinfo {title} {Solar wind turbulence and
  the role of ion instabilities},}\ }\href@noop {} {\bibfield  {journal}
  {\bibinfo  {journal} {Space Science Reviews}\ }\textbf {\bibinfo {volume}
  {178}},\ \bibinfo {pages} {101--139} (\bibinfo {year} {2013})}\BibitemShut
  {NoStop}%
\bibitem [{\citenamefont {Andr\'es}\ and\ \citenamefont
  {Banerjee}(2019)}]{A2019}%
  \BibitemOpen
  \bibfield  {author} {\bibinfo {author} {\bibnamefont {Andr\'es},
  \bibfnamefont {N.}}\ and\ \bibinfo {author} {\bibnamefont {Banerjee},
  \bibfnamefont {S.}},\ }\bibfield  {title} {\enquote {\bibinfo {title}
  {Statistics of incompressible hydrodynamic turbulence: An alternative
  approach},}\ }\href {\doibase 10.1103/PhysRevFluids.4.024603} {\bibfield
  {journal} {\bibinfo  {journal} {Phys. Rev. Fluids}\ }\textbf {\bibinfo
  {volume} {4}},\ \bibinfo {pages} {024603} (\bibinfo {year}
  {2019})}\BibitemShut {NoStop}%
\bibitem [{\citenamefont {Andr{\'e}s}, \citenamefont {Galtier},\ and\
  \citenamefont {Sahraoui}(2018)}]{A2018}%
  \BibitemOpen
  \bibfield  {author} {\bibinfo {author} {\bibnamefont {Andr{\'e}s},
  \bibfnamefont {N.}}, \bibinfo {author} {\bibnamefont {Galtier}, \bibfnamefont
  {S.}}, \ and\ \bibinfo {author} {\bibnamefont {Sahraoui}, \bibfnamefont
  {F.}},\ }\bibfield  {title} {\enquote {\bibinfo {title} {Exact law for
  homogeneous compressible hall magnetohydrodynamics turbulence},}\ }\href@noop
  {} {\bibfield  {journal} {\bibinfo  {journal} {Physical Review E}\ }\textbf
  {\bibinfo {volume} {97}},\ \bibinfo {pages} {013204} (\bibinfo {year}
  {2018})}\BibitemShut {NoStop}%
\bibitem [{\citenamefont {Andr{\'e}s}\ \emph {et~al.}(2020)\citenamefont
  {Andr{\'e}s}, \citenamefont {Romanelli}, \citenamefont {Hadid}, \citenamefont
  {Sahraoui}, \citenamefont {DiBraccio},\ and\ \citenamefont
  {Halekas}}]{A2020}%
  \BibitemOpen
  \bibfield  {author} {\bibinfo {author} {\bibnamefont {Andr{\'e}s},
  \bibfnamefont {N.}}, \bibinfo {author} {\bibnamefont {Romanelli},
  \bibfnamefont {N.}}, \bibinfo {author} {\bibnamefont {Hadid}, \bibfnamefont
  {L.~Z.}}, \bibinfo {author} {\bibnamefont {Sahraoui}, \bibfnamefont {F.}},
  \bibinfo {author} {\bibnamefont {DiBraccio}, \bibfnamefont {G.}}, \ and\
  \bibinfo {author} {\bibnamefont {Halekas}, \bibfnamefont {J.}},\ }\bibfield
  {title} {\enquote {\bibinfo {title} {Solar wind turbulence around mars:
  Relation between the energy cascade rate and the proton cyclotron waves
  activity},}\ }\href@noop {} {\bibfield  {journal} {\bibinfo  {journal} {The
  Astrophysical Journal}\ }\textbf {\bibinfo {volume} {902}},\ \bibinfo {pages}
  {134} (\bibinfo {year} {2020})}\BibitemShut {NoStop}%
\bibitem [{\citenamefont {Andr{\'e}s}\ and\ \citenamefont
  {Sahraoui}(2017)}]{A2017b}%
  \BibitemOpen
  \bibfield  {author} {\bibinfo {author} {\bibnamefont {Andr{\'e}s},
  \bibfnamefont {N.}}\ and\ \bibinfo {author} {\bibnamefont {Sahraoui},
  \bibfnamefont {F.}},\ }\bibfield  {title} {\enquote {\bibinfo {title}
  {Alternative derivation of exact law for compressible and isothermal
  magnetohydrodynamics turbulence},}\ }\href@noop {} {\bibfield  {journal}
  {\bibinfo  {journal} {Physical Review E}\ }\textbf {\bibinfo {volume} {96}},\
  \bibinfo {pages} {053205} (\bibinfo {year} {2017})}\BibitemShut {NoStop}%
\bibitem [{\citenamefont {Andr\'es}\ \emph {et~al.}(2019)\citenamefont
  {Andr\'es}, \citenamefont {Sahraoui}, \citenamefont {Galtier}, \citenamefont
  {Hadid}, \citenamefont {Ferrand},\ and\ \citenamefont {Huang}}]{A2019b}%
  \BibitemOpen
  \bibfield  {author} {\bibinfo {author} {\bibnamefont {Andr\'es},
  \bibfnamefont {N.}}, \bibinfo {author} {\bibnamefont {Sahraoui},
  \bibfnamefont {F.}}, \bibinfo {author} {\bibnamefont {Galtier}, \bibfnamefont
  {S.}}, \bibinfo {author} {\bibnamefont {Hadid}, \bibfnamefont {L.~Z.}},
  \bibinfo {author} {\bibnamefont {Ferrand}, \bibfnamefont {R.}}, \ and\
  \bibinfo {author} {\bibnamefont {Huang}, \bibfnamefont {S.~Y.}},\ }\bibfield
  {title} {\enquote {\bibinfo {title} {Energy cascade rate measured in a
  collisionless space plasma with mms data and compressible hall
  magnetohydrodynamic turbulence theory},}\ }\href {\doibase
  10.1103/PhysRevLett.123.245101} {\bibfield  {journal} {\bibinfo  {journal}
  {Phys. Rev. Lett.}\ }\textbf {\bibinfo {volume} {123}},\ \bibinfo {pages}
  {245101} (\bibinfo {year} {2019})}\BibitemShut {NoStop}%
\bibitem [{\citenamefont {Andr\'es}\ \emph {et~al.}(2021)\citenamefont
  {Andr\'es}, \citenamefont {Sahraoui}, \citenamefont {Hadid}, \citenamefont
  {Huang}, \citenamefont {Romanelli}, \citenamefont {Galtier}, \citenamefont
  {DiBraccio},\ and\ \citenamefont {Halekas}}]{A2021}%
  \BibitemOpen
  \bibfield  {author} {\bibinfo {author} {\bibnamefont {Andr\'es},
  \bibfnamefont {N.}}, \bibinfo {author} {\bibnamefont {Sahraoui},
  \bibfnamefont {F.}}, \bibinfo {author} {\bibnamefont {Hadid}, \bibfnamefont
  {L.~Z.}}, \bibinfo {author} {\bibnamefont {Huang}, \bibfnamefont {S.~Y.}},
  \bibinfo {author} {\bibnamefont {Romanelli}, \bibfnamefont {N.}}, \bibinfo
  {author} {\bibnamefont {Galtier}, \bibfnamefont {S.}}, \bibinfo {author}
  {\bibnamefont {DiBraccio}, \bibfnamefont {G.}}, \ and\ \bibinfo {author}
  {\bibnamefont {Halekas}, \bibfnamefont {J.}},\ }\bibfield  {title} {\enquote
  {\bibinfo {title} {The evolution of compressible solar wind turbulence in the
  inner heliosphere: Psp,themis and maven observations},}\ }\href@noop {}
  {\bibfield  {journal} {\bibinfo  {journal} {In press ApJ}\ } (\bibinfo {year}
  {2021})}\BibitemShut {NoStop}%
\bibitem [{\citenamefont {Andr{\'e}s}\ \emph {et~al.}(2022)\citenamefont
  {Andr{\'e}s}, \citenamefont {Sahraoui}, \citenamefont {Huang}, \citenamefont
  {Hadid},\ and\ \citenamefont {Galtier}}]{A2022}%
  \BibitemOpen
  \bibfield  {author} {\bibinfo {author} {\bibnamefont {Andr{\'e}s},
  \bibfnamefont {N.}}, \bibinfo {author} {\bibnamefont {Sahraoui},
  \bibfnamefont {F.}}, \bibinfo {author} {\bibnamefont {Huang}, \bibfnamefont
  {S.}}, \bibinfo {author} {\bibnamefont {Hadid}, \bibfnamefont {L.}}, \ and\
  \bibinfo {author} {\bibnamefont {Galtier}, \bibfnamefont {S.}},\ }\bibfield
  {title} {\enquote {\bibinfo {title} {The incompressible energy cascade rate
  in anisotropic solar wind turbulence},}\ }\href@noop {} {\bibfield  {journal}
  {\bibinfo  {journal} {Astronomy \& Astrophysics}\ }\textbf {\bibinfo {volume}
  {661}},\ \bibinfo {pages} {A116} (\bibinfo {year} {2022})}\BibitemShut
  {NoStop}%
\bibitem [{\citenamefont {Auster}\ \emph {et~al.}(2009)\citenamefont {Auster},
  \citenamefont {Glassmeier}, \citenamefont {Magnes}, \citenamefont {Aydogar},
  \citenamefont {Baumjohann}, \citenamefont {Constantinescu}, \citenamefont
  {Fischer}, \citenamefont {Fornacon}, \citenamefont {Georgescu}, \citenamefont
  {Harvey} \emph {et~al.}}]{Au2009}%
  \BibitemOpen
  \bibfield  {author} {\bibinfo {author} {\bibnamefont {Auster}, \bibfnamefont
  {H.~U.}}, \bibinfo {author} {\bibnamefont {Glassmeier}, \bibfnamefont
  {K.~H.}}, \bibinfo {author} {\bibnamefont {Magnes}, \bibfnamefont {W.}},
  \bibinfo {author} {\bibnamefont {Aydogar}, \bibfnamefont {O.}}, \bibinfo
  {author} {\bibnamefont {Baumjohann}, \bibfnamefont {W.}}, \bibinfo {author}
  {\bibnamefont {Constantinescu}, \bibfnamefont {D.}}, \bibinfo {author}
  {\bibnamefont {Fischer}, \bibfnamefont {D.}}, \bibinfo {author} {\bibnamefont
  {Fornacon}, \bibfnamefont {K.~H.}}, \bibinfo {author} {\bibnamefont
  {Georgescu}, \bibfnamefont {E.}}, \bibinfo {author} {\bibnamefont {Harvey},
  \bibfnamefont {P.}},  \emph {et~al.},\ }\bibfield  {title} {\enquote
  {\bibinfo {title} {The themis fluxgate magnetometer},}\ }in\ \href@noop {}
  {\emph {\bibinfo {booktitle} {The THEMIS Mission}}}\ (\bibinfo  {publisher}
  {Springer},\ \bibinfo {year} {2009})\ pp.\ \bibinfo {pages}
  {235--264}\BibitemShut {NoStop}%
\bibitem [{\citenamefont {Bale}\ \emph {et~al.}(2019)\citenamefont {Bale},
  \citenamefont {Badman}, \citenamefont {Bonnell}, \citenamefont {Bowen},
  \citenamefont {Burgess}, \citenamefont {Case}, \citenamefont {Cattell},
  \citenamefont {Chandran}, \citenamefont {Chaston}, \citenamefont {Chen} \emph
  {et~al.}}]{B2019}%
  \BibitemOpen
  \bibfield  {author} {\bibinfo {author} {\bibnamefont {Bale}, \bibfnamefont
  {S.}}, \bibinfo {author} {\bibnamefont {Badman}, \bibfnamefont {S.}},
  \bibinfo {author} {\bibnamefont {Bonnell}, \bibfnamefont {J.}}, \bibinfo
  {author} {\bibnamefont {Bowen}, \bibfnamefont {T.}}, \bibinfo {author}
  {\bibnamefont {Burgess}, \bibfnamefont {D.}}, \bibinfo {author} {\bibnamefont
  {Case}, \bibfnamefont {A.}}, \bibinfo {author} {\bibnamefont {Cattell},
  \bibfnamefont {C.}}, \bibinfo {author} {\bibnamefont {Chandran},
  \bibfnamefont {B.}}, \bibinfo {author} {\bibnamefont {Chaston}, \bibfnamefont
  {C.}}, \bibinfo {author} {\bibnamefont {Chen}, \bibfnamefont {C.}},  \emph
  {et~al.},\ }\bibfield  {title} {\enquote {\bibinfo {title} {Highly structured
  slow solar wind emerging from an equatorial coronal hole},}\ }\href@noop {}
  {\bibfield  {journal} {\bibinfo  {journal} {Nature}\ }\textbf {\bibinfo
  {volume} {576}},\ \bibinfo {pages} {237--242} (\bibinfo {year}
  {2019})}\BibitemShut {NoStop}%
\bibitem [{\citenamefont {Bale}\ \emph {et~al.}(2016)\citenamefont {Bale},
  \citenamefont {Goetz}, \citenamefont {Harvey}, \citenamefont {Turin},
  \citenamefont {Bonnell}, \citenamefont {De~Wit}, \citenamefont {Ergun},
  \citenamefont {MacDowall}, \citenamefont {Pulupa}, \citenamefont {Andr{\'e}}
  \emph {et~al.}}]{Ba2016}%
  \BibitemOpen
  \bibfield  {author} {\bibinfo {author} {\bibnamefont {Bale}, \bibfnamefont
  {S.}}, \bibinfo {author} {\bibnamefont {Goetz}, \bibfnamefont {K.}}, \bibinfo
  {author} {\bibnamefont {Harvey}, \bibfnamefont {P.}}, \bibinfo {author}
  {\bibnamefont {Turin}, \bibfnamefont {P.}}, \bibinfo {author} {\bibnamefont
  {Bonnell}, \bibfnamefont {J.}}, \bibinfo {author} {\bibnamefont {De~Wit},
  \bibfnamefont {T.~D.}}, \bibinfo {author} {\bibnamefont {Ergun},
  \bibfnamefont {R.}}, \bibinfo {author} {\bibnamefont {MacDowall},
  \bibfnamefont {R.}}, \bibinfo {author} {\bibnamefont {Pulupa}, \bibfnamefont
  {M.}}, \bibinfo {author} {\bibnamefont {Andr{\'e}}, \bibfnamefont {M.}},
  \emph {et~al.},\ }\bibfield  {title} {\enquote {\bibinfo {title} {The fields
  instrument suite for solar probe plus},}\ }\href@noop {} {\bibfield
  {journal} {\bibinfo  {journal} {Space science reviews}\ }\textbf {\bibinfo
  {volume} {204}},\ \bibinfo {pages} {49--82} (\bibinfo {year}
  {2016})}\BibitemShut {NoStop}%
\bibitem [{\citenamefont {Bandyopadhyay}\ \emph {et~al.}(2018)\citenamefont
  {Bandyopadhyay}, \citenamefont {Chasapis}, \citenamefont {Chhiber},
  \citenamefont {Parashar}, \citenamefont {Maruca}, \citenamefont {Matthaeus},
  \citenamefont {Schwartz}, \citenamefont {Eriksson}, \citenamefont {Contel},
  \citenamefont {Breuillard}, \citenamefont {Burch}, \citenamefont {Moore},
  \citenamefont {Pollock}, \citenamefont {Giles}, \citenamefont {Paterson},
  \citenamefont {Dorelli}, \citenamefont {Gershman}, \citenamefont {Torbert},
  \citenamefont {Russell},\ and\ \citenamefont {Strangeway}}]{riddhi2018}%
  \BibitemOpen
  \bibfield  {author} {\bibinfo {author} {\bibnamefont {Bandyopadhyay},
  \bibfnamefont {R.}}, \bibinfo {author} {\bibnamefont {Chasapis},
  \bibfnamefont {A.}}, \bibinfo {author} {\bibnamefont {Chhiber}, \bibfnamefont
  {R.}}, \bibinfo {author} {\bibnamefont {Parashar}, \bibfnamefont {T.~N.}},
  \bibinfo {author} {\bibnamefont {Maruca}, \bibfnamefont {B.~A.}}, \bibinfo
  {author} {\bibnamefont {Matthaeus}, \bibfnamefont {W.~H.}}, \bibinfo {author}
  {\bibnamefont {Schwartz}, \bibfnamefont {S.~J.}}, \bibinfo {author}
  {\bibnamefont {Eriksson}, \bibfnamefont {S.}}, \bibinfo {author}
  {\bibnamefont {Contel}, \bibfnamefont {O.~L.}}, \bibinfo {author}
  {\bibnamefont {Breuillard}, \bibfnamefont {H.}}, \bibinfo {author}
  {\bibnamefont {Burch}, \bibfnamefont {J.~L.}}, \bibinfo {author}
  {\bibnamefont {Moore}, \bibfnamefont {T.~E.}}, \bibinfo {author}
  {\bibnamefont {Pollock}, \bibfnamefont {C.~J.}}, \bibinfo {author}
  {\bibnamefont {Giles}, \bibfnamefont {B.~L.}}, \bibinfo {author}
  {\bibnamefont {Paterson}, \bibfnamefont {W.~R.}}, \bibinfo {author}
  {\bibnamefont {Dorelli}, \bibfnamefont {J.}}, \bibinfo {author} {\bibnamefont
  {Gershman}, \bibfnamefont {D.~J.}}, \bibinfo {author} {\bibnamefont
  {Torbert}, \bibfnamefont {R.~B.}}, \bibinfo {author} {\bibnamefont {Russell},
  \bibfnamefont {C.~T.}}, \ and\ \bibinfo {author} {\bibnamefont {Strangeway},
  \bibfnamefont {R.~J.}},\ }\bibfield  {title} {\enquote {\bibinfo {title}
  {Solar wind turbulence studies using {MMS} fast plasma investigation data},}\
  }\href {\doibase 10.3847/1538-4357/aade93} {\bibfield  {journal} {\bibinfo
  {journal} {The Astrophysical Journal}\ }\textbf {\bibinfo {volume} {866}},\
  \bibinfo {pages} {81} (\bibinfo {year} {2018})}\BibitemShut {NoStop}%
\bibitem [{\citenamefont {Bandyopadhyay}\ \emph {et~al.}(2020)\citenamefont
  {Bandyopadhyay}, \citenamefont {Goldstein}, \citenamefont {Maruca},
  \citenamefont {Matthaeus}, \citenamefont {Parashar}, \citenamefont {Ruffolo},
  \citenamefont {Chhiber}, \citenamefont {Usmanov}, \citenamefont {Chasapis},
  \citenamefont {Qudsi} \emph {et~al.}}]{B2020}%
  \BibitemOpen
  \bibfield  {author} {\bibinfo {author} {\bibnamefont {Bandyopadhyay},
  \bibfnamefont {R.}}, \bibinfo {author} {\bibnamefont {Goldstein},
  \bibfnamefont {M.}}, \bibinfo {author} {\bibnamefont {Maruca}, \bibfnamefont
  {B.}}, \bibinfo {author} {\bibnamefont {Matthaeus}, \bibfnamefont {W.}},
  \bibinfo {author} {\bibnamefont {Parashar}, \bibfnamefont {T.}}, \bibinfo
  {author} {\bibnamefont {Ruffolo}, \bibfnamefont {D.}}, \bibinfo {author}
  {\bibnamefont {Chhiber}, \bibfnamefont {R.}}, \bibinfo {author} {\bibnamefont
  {Usmanov}, \bibfnamefont {A.}}, \bibinfo {author} {\bibnamefont {Chasapis},
  \bibfnamefont {A.}}, \bibinfo {author} {\bibnamefont {Qudsi}, \bibfnamefont
  {R.}},  \emph {et~al.},\ }\bibfield  {title} {\enquote {\bibinfo {title}
  {Enhanced energy transfer rate in solar wind turbulence observed near the sun
  from parker solar probe},}\ }\href@noop {} {\bibfield  {journal} {\bibinfo
  {journal} {The Astrophysical Journal Supplement Series}\ }\textbf {\bibinfo
  {volume} {246}},\ \bibinfo {pages} {48} (\bibinfo {year} {2020})}\BibitemShut
  {NoStop}%
\bibitem [{\citenamefont {Banerjee}\ and\ \citenamefont
  {Andr{\'e}s}(2020)}]{Ba2020}%
  \BibitemOpen
  \bibfield  {author} {\bibinfo {author} {\bibnamefont {Banerjee},
  \bibfnamefont {S.}}\ and\ \bibinfo {author} {\bibnamefont {Andr{\'e}s},
  \bibfnamefont {N.}},\ }\bibfield  {title} {\enquote {\bibinfo {title}
  {Scale-to-scale energy transfer rate in compressible two-fluid plasma
  turbulence},}\ }\href@noop {} {\bibfield  {journal} {\bibinfo  {journal}
  {Physical Review E}\ }\textbf {\bibinfo {volume} {101}},\ \bibinfo {pages}
  {043212} (\bibinfo {year} {2020})}\BibitemShut {NoStop}%
\bibitem [{\citenamefont {Banerjee}\ and\ \citenamefont
  {Galtier}(2013)}]{B2013}%
  \BibitemOpen
  \bibfield  {author} {\bibinfo {author} {\bibnamefont {Banerjee},
  \bibfnamefont {S.}}\ and\ \bibinfo {author} {\bibnamefont {Galtier},
  \bibfnamefont {S.}},\ }\bibfield  {title} {\enquote {\bibinfo {title} {Exact
  relation with two-point correlation functions and phenomenological approach
  for compressible magnetohydrodynamic turbulence},}\ }\href@noop {} {\bibfield
   {journal} {\bibinfo  {journal} {Physical Review E}\ }\textbf {\bibinfo
  {volume} {87}},\ \bibinfo {pages} {013019} (\bibinfo {year}
  {2013})}\BibitemShut {NoStop}%
\bibitem [{\citenamefont {Banerjee}\ and\ \citenamefont
  {Galtier}(2014)}]{B2014}%
  \BibitemOpen
  \bibfield  {author} {\bibinfo {author} {\bibnamefont {Banerjee},
  \bibfnamefont {S.}}\ and\ \bibinfo {author} {\bibnamefont {Galtier},
  \bibfnamefont {S.}},\ }\bibfield  {title} {\enquote {\bibinfo {title} {A
  kolmogorov-like exact relation for compressible polytropic turbulence},}\
  }\href@noop {} {\bibfield  {journal} {\bibinfo  {journal} {Journal of Fluid
  Mechanics}\ }\textbf {\bibinfo {volume} {742}},\ \bibinfo {pages} {230--242}
  (\bibinfo {year} {2014})}\BibitemShut {NoStop}%
\bibitem [{\citenamefont {Banerjee}\ \emph {et~al.}(2016)\citenamefont
  {Banerjee}, \citenamefont {Hadid}, \citenamefont {Sahraoui},\ and\
  \citenamefont {Galtier}}]{B2016c}%
  \BibitemOpen
  \bibfield  {author} {\bibinfo {author} {\bibnamefont {Banerjee},
  \bibfnamefont {S.}}, \bibinfo {author} {\bibnamefont {Hadid}, \bibfnamefont
  {L.~Z.}}, \bibinfo {author} {\bibnamefont {Sahraoui}, \bibfnamefont {F.}}, \
  and\ \bibinfo {author} {\bibnamefont {Galtier}, \bibfnamefont {S.}},\
  }\bibfield  {title} {\enquote {\bibinfo {title} {Scaling of compressible
  magnetohydrodynamic turbulence in the fast solar wind},}\ }\href@noop {}
  {\bibfield  {journal} {\bibinfo  {journal} {The Astrophysical Journal
  Letters}\ }\textbf {\bibinfo {volume} {829}},\ \bibinfo {pages} {L27}
  (\bibinfo {year} {2016})}\BibitemShut {NoStop}%
\bibitem [{\citenamefont {Banerjee}\ and\ \citenamefont
  {Kritsuk}(2018)}]{B2018}%
  \BibitemOpen
  \bibfield  {author} {\bibinfo {author} {\bibnamefont {Banerjee},
  \bibfnamefont {S.}}\ and\ \bibinfo {author} {\bibnamefont {Kritsuk},
  \bibfnamefont {A.~G.}},\ }\bibfield  {title} {\enquote {\bibinfo {title}
  {Energy transfer in compressible magnetohydrodynamic turbulence for
  isothermal self-gravitating fluids},}\ }\href {\doibase
  10.1103/PhysRevE.97.023107} {\bibfield  {journal} {\bibinfo  {journal} {Phys.
  Rev. E}\ }\textbf {\bibinfo {volume} {97}},\ \bibinfo {pages} {023107}
  (\bibinfo {year} {2018})}\BibitemShut {NoStop}%
\bibitem [{\citenamefont {Boldyrev}, \citenamefont {Mason},\ and\ \citenamefont
  {Cattaneo}(2009)}]{Bo2009}%
  \BibitemOpen
  \bibfield  {author} {\bibinfo {author} {\bibnamefont {Boldyrev},
  \bibfnamefont {S.}}, \bibinfo {author} {\bibnamefont {Mason}, \bibfnamefont
  {J.}}, \ and\ \bibinfo {author} {\bibnamefont {Cattaneo}, \bibfnamefont
  {F.}},\ }\bibfield  {title} {\enquote {\bibinfo {title} {Dynamic alignment
  and exact scaling laws in magnetohydrodynamic turbulence},}\ }\href@noop {}
  {\bibfield  {journal} {\bibinfo  {journal} {Astrophys. J. Lett.}\ }\textbf
  {\bibinfo {volume} {699}},\ \bibinfo {pages} {L39} (\bibinfo {year}
  {2009})}\BibitemShut {NoStop}%
\bibitem [{\citenamefont {Bourouaine}\ \emph {et~al.}(2020)\citenamefont
  {Bourouaine}, \citenamefont {Perez}, \citenamefont {Klein}, \citenamefont
  {Chen}, \citenamefont {Martinović}, \citenamefont {Bale}, \citenamefont
  {Kasper},\ and\ \citenamefont {Raouafi}}]{Bour2020}%
  \BibitemOpen
  \bibfield  {author} {\bibinfo {author} {\bibnamefont {Bourouaine},
  \bibfnamefont {S.}}, \bibinfo {author} {\bibnamefont {Perez}, \bibfnamefont
  {J.~C.}}, \bibinfo {author} {\bibnamefont {Klein}, \bibfnamefont {K.~G.}},
  \bibinfo {author} {\bibnamefont {Chen}, \bibfnamefont {C.~H.~K.}}, \bibinfo
  {author} {\bibnamefont {Martinović}, \bibfnamefont {M.}}, \bibinfo {author}
  {\bibnamefont {Bale}, \bibfnamefont {S.~D.}}, \bibinfo {author} {\bibnamefont
  {Kasper}, \bibfnamefont {J.~C.}}, \ and\ \bibinfo {author} {\bibnamefont
  {Raouafi}, \bibfnamefont {N.~E.}},\ }\bibfield  {title} {\enquote {\bibinfo
  {title} {Turbulence characteristics of switchback and nonswitchback intervals
  observed by parker solar probe},}\ }\href {\doibase
  10.3847/2041-8213/abbd4a} {\bibfield  {journal} {\bibinfo  {journal} {The
  Astrophysical Journal Letters}\ }\textbf {\bibinfo {volume} {904}},\ \bibinfo
  {pages} {L30} (\bibinfo {year} {2020})}\BibitemShut {NoStop}%
\bibitem [{\citenamefont {Bruno}\ and\ \citenamefont
  {Carbone}(2013)}]{bruno2013}%
  \BibitemOpen
  \bibfield  {author} {\bibinfo {author} {\bibnamefont {Bruno}, \bibfnamefont
  {R.}}\ and\ \bibinfo {author} {\bibnamefont {Carbone}, \bibfnamefont {V.}},\
  }\bibfield  {title} {\enquote {\bibinfo {title} {The solar wind as a
  turbulence laboratory},}\ }\href@noop {} {\bibfield  {journal} {\bibinfo
  {journal} {Living Reviews in Solar Physics}\ }\textbf {\bibinfo {volume}
  {10}},\ \bibinfo {pages} {1--208} (\bibinfo {year} {2013})}\BibitemShut
  {NoStop}%
\bibitem [{\citenamefont {Carbone}\ \emph {et~al.}(2009)\citenamefont
  {Carbone}, \citenamefont {Marino}, \citenamefont {Sorriso-Valvo},
  \citenamefont {Noullez},\ and\ \citenamefont {Bruno}}]{C2009b}%
  \BibitemOpen
  \bibfield  {author} {\bibinfo {author} {\bibnamefont {Carbone}, \bibfnamefont
  {V.}}, \bibinfo {author} {\bibnamefont {Marino}, \bibfnamefont {R.}},
  \bibinfo {author} {\bibnamefont {Sorriso-Valvo}, \bibfnamefont {L.}},
  \bibinfo {author} {\bibnamefont {Noullez}, \bibfnamefont {A.}}, \ and\
  \bibinfo {author} {\bibnamefont {Bruno}, \bibfnamefont {R.}},\ }\bibfield
  {title} {\enquote {\bibinfo {title} {Scaling laws of turbulence and heating
  of fast solar wind: the role of density fluctuations},}\ }\href@noop {}
  {\bibfield  {journal} {\bibinfo  {journal} {Physical review letters}\
  }\textbf {\bibinfo {volume} {103}},\ \bibinfo {pages} {061102} (\bibinfo
  {year} {2009})}\BibitemShut {NoStop}%
\bibitem [{\citenamefont {Case}\ \emph {et~al.}(2020)\citenamefont {Case},
  \citenamefont {Kasper}, \citenamefont {Stevens}, \citenamefont {Korreck},
  \citenamefont {Paulson}, \citenamefont {Daigneau}, \citenamefont {Caldwell},
  \citenamefont {Freeman}, \citenamefont {Henry}, \citenamefont {Klingensmith}
  \emph {et~al.}}]{C2020}%
  \BibitemOpen
  \bibfield  {author} {\bibinfo {author} {\bibnamefont {Case}, \bibfnamefont
  {A.~W.}}, \bibinfo {author} {\bibnamefont {Kasper}, \bibfnamefont {J.~C.}},
  \bibinfo {author} {\bibnamefont {Stevens}, \bibfnamefont {M.~L.}}, \bibinfo
  {author} {\bibnamefont {Korreck}, \bibfnamefont {K.~E.}}, \bibinfo {author}
  {\bibnamefont {Paulson}, \bibfnamefont {K.}}, \bibinfo {author} {\bibnamefont
  {Daigneau}, \bibfnamefont {P.}}, \bibinfo {author} {\bibnamefont {Caldwell},
  \bibfnamefont {D.}}, \bibinfo {author} {\bibnamefont {Freeman}, \bibfnamefont
  {M.}}, \bibinfo {author} {\bibnamefont {Henry}, \bibfnamefont {T.}}, \bibinfo
  {author} {\bibnamefont {Klingensmith}, \bibfnamefont {B.}},  \emph {et~al.},\
  }\bibfield  {title} {\enquote {\bibinfo {title} {The solar probe cup on the
  parker solar probe},}\ }\href@noop {} {\bibfield  {journal} {\bibinfo
  {journal} {The Astrophysical Journal Supplement Series}\ }\textbf {\bibinfo
  {volume} {246}},\ \bibinfo {pages} {43} (\bibinfo {year} {2020})}\BibitemShut
  {NoStop}%
\bibitem [{\citenamefont {Chen}\ \emph {et~al.}(2020)\citenamefont {Chen},
  \citenamefont {Bale}, \citenamefont {Bonnell}, \citenamefont {Borovikov},
  \citenamefont {Bowen}, \citenamefont {Burgess}, \citenamefont {Case},
  \citenamefont {Chandran}, \citenamefont {de~Wit}, \citenamefont {Goetz} \emph
  {et~al.}}]{Ch2020}%
  \BibitemOpen
  \bibfield  {author} {\bibinfo {author} {\bibnamefont {Chen}, \bibfnamefont
  {C.}}, \bibinfo {author} {\bibnamefont {Bale}, \bibfnamefont {S.}}, \bibinfo
  {author} {\bibnamefont {Bonnell}, \bibfnamefont {J.}}, \bibinfo {author}
  {\bibnamefont {Borovikov}, \bibfnamefont {D.}}, \bibinfo {author}
  {\bibnamefont {Bowen}, \bibfnamefont {T.}}, \bibinfo {author} {\bibnamefont
  {Burgess}, \bibfnamefont {D.}}, \bibinfo {author} {\bibnamefont {Case},
  \bibfnamefont {A.}}, \bibinfo {author} {\bibnamefont {Chandran},
  \bibfnamefont {B.}}, \bibinfo {author} {\bibnamefont {de~Wit}, \bibfnamefont
  {T.~D.}}, \bibinfo {author} {\bibnamefont {Goetz}, \bibfnamefont {K.}},
  \emph {et~al.},\ }\bibfield  {title} {\enquote {\bibinfo {title} {The
  evolution and role of solar wind turbulence in the inner heliosphere},}\
  }\href@noop {} {\bibfield  {journal} {\bibinfo  {journal} {The Astrophysical
  Journal Supplement Series}\ }\textbf {\bibinfo {volume} {246}},\ \bibinfo
  {pages} {53} (\bibinfo {year} {2020})}\BibitemShut {NoStop}%
\bibitem [{\citenamefont {Chen}(2016)}]{Ch2016}%
  \BibitemOpen
  \bibfield  {author} {\bibinfo {author} {\bibnamefont {Chen}, \bibfnamefont
  {C.~H.~K.}},\ }\bibfield  {title} {\enquote {\bibinfo {title} {Recent
  progress in astrophysical plasma turbulence from solar wind observations},}\
  }\href {\doibase 10.1017/S0022377816001124} {\bibfield  {journal} {\bibinfo
  {journal} {Journal of Plasma Physics}\ }\textbf {\bibinfo {volume} {82}},\
  \bibinfo {pages} {535820602} (\bibinfo {year} {2016})}\BibitemShut {NoStop}%
\bibitem [{\citenamefont {Coburn}\ \emph {et~al.}(2015)\citenamefont {Coburn},
  \citenamefont {Forman}, \citenamefont {Smith}, \citenamefont {Vasquez},\ and\
  \citenamefont {Stawarz}}]{Co2015}%
  \BibitemOpen
  \bibfield  {author} {\bibinfo {author} {\bibnamefont {Coburn}, \bibfnamefont
  {J.~T.}}, \bibinfo {author} {\bibnamefont {Forman}, \bibfnamefont {M.~A.}},
  \bibinfo {author} {\bibnamefont {Smith}, \bibfnamefont {C.~W.}}, \bibinfo
  {author} {\bibnamefont {Vasquez}, \bibfnamefont {B.~J.}}, \ and\ \bibinfo
  {author} {\bibnamefont {Stawarz}, \bibfnamefont {J.~E.}},\ }\bibfield
  {title} {\enquote {\bibinfo {title} {Third-moment descriptions of the
  interplanetary turbulent cascade, intermittency and back transfer},}\ }\href
  {\doibase 10.1098/rsta.2014.0150} {\bibfield  {journal} {\bibinfo  {journal}
  {Philosophical Transactions of the Royal Society A: Mathematical, Physical
  and Engineering Sciences}\ }\textbf {\bibinfo {volume} {373}},\ \bibinfo
  {pages} {20140150} (\bibinfo {year} {2015})}\BibitemShut {NoStop}%
\bibitem [{\citenamefont {Davies}\ and\ \citenamefont
  {Gather}(1993)}]{Davies1993}%
  \BibitemOpen
  \bibfield  {author} {\bibinfo {author} {\bibnamefont {Davies}, \bibfnamefont
  {P.}}\ and\ \bibinfo {author} {\bibnamefont {Gather}, \bibfnamefont {U.}},\
  }\bibfield  {title} {{\selectlanguage {English}\enquote {\bibinfo {title}
  {The identification of multiple outliers},}\ }}\href {\doibase
  10.2307/2290763} {\bibfield  {journal} {\bibinfo  {journal} {Journal of the
  American Statistical Association}\ }\textbf {\bibinfo {volume} {88}},\
  \bibinfo {pages} {782--792} (\bibinfo {year} {1993})}\BibitemShut {NoStop}%
\bibitem [{\citenamefont {Ferrand}\ \emph {et~al.}(2020)\citenamefont
  {Ferrand}, \citenamefont {Galtier}, \citenamefont {Sahraoui},\ and\
  \citenamefont {Federrath}}]{F2020}%
  \BibitemOpen
  \bibfield  {author} {\bibinfo {author} {\bibnamefont {Ferrand}, \bibfnamefont
  {R.}}, \bibinfo {author} {\bibnamefont {Galtier}, \bibfnamefont {S.}},
  \bibinfo {author} {\bibnamefont {Sahraoui}, \bibfnamefont {F.}}, \ and\
  \bibinfo {author} {\bibnamefont {Federrath}, \bibfnamefont {C.}},\ }\bibfield
   {title} {\enquote {\bibinfo {title} {Compressible turbulence in the
  interstellar medium: New insights from a high-resolution supersonic
  turbulence simulation},}\ }\href {\doibase 10.3847/1538-4357/abb76e}
  {\bibfield  {journal} {\bibinfo  {journal} {The Astrophysical Journal}\
  }\textbf {\bibinfo {volume} {904}},\ \bibinfo {pages} {160} (\bibinfo {year}
  {2020})}\BibitemShut {NoStop}%
\bibitem [{\citenamefont {Fox}\ \emph {et~al.}(2016)\citenamefont {Fox},
  \citenamefont {Velli}, \citenamefont {Bale}, \citenamefont {Decker},
  \citenamefont {Driesman}, \citenamefont {Howard}, \citenamefont {Kasper},
  \citenamefont {Kinnison}, \citenamefont {Kusterer}, \citenamefont {Lario}
  \emph {et~al.}}]{Fo2016}%
  \BibitemOpen
  \bibfield  {author} {\bibinfo {author} {\bibnamefont {Fox}, \bibfnamefont
  {N.}}, \bibinfo {author} {\bibnamefont {Velli}, \bibfnamefont {M.}}, \bibinfo
  {author} {\bibnamefont {Bale}, \bibfnamefont {S.}}, \bibinfo {author}
  {\bibnamefont {Decker}, \bibfnamefont {R.}}, \bibinfo {author} {\bibnamefont
  {Driesman}, \bibfnamefont {A.}}, \bibinfo {author} {\bibnamefont {Howard},
  \bibfnamefont {R.}}, \bibinfo {author} {\bibnamefont {Kasper}, \bibfnamefont
  {J.~C.}}, \bibinfo {author} {\bibnamefont {Kinnison}, \bibfnamefont {J.}},
  \bibinfo {author} {\bibnamefont {Kusterer}, \bibfnamefont {M.}}, \bibinfo
  {author} {\bibnamefont {Lario}, \bibfnamefont {D.}},  \emph {et~al.},\
  }\bibfield  {title} {\enquote {\bibinfo {title} {The solar probe plus
  mission: humanity’s first visit to our star},}\ }\href@noop {} {\bibfield
  {journal} {\bibinfo  {journal} {Space Science Reviews}\ }\textbf {\bibinfo
  {volume} {204}},\ \bibinfo {pages} {7--48} (\bibinfo {year}
  {2016})}\BibitemShut {NoStop}%
\bibitem [{\citenamefont {Fraternale}\ \emph {et~al.}(2019)\citenamefont
  {Fraternale}, \citenamefont {Pogorelov}, \citenamefont {Richardson},\ and\
  \citenamefont {Tordella}}]{Fra2019}%
  \BibitemOpen
  \bibfield  {author} {\bibinfo {author} {\bibnamefont {Fraternale},
  \bibfnamefont {F.}}, \bibinfo {author} {\bibnamefont {Pogorelov},
  \bibfnamefont {N.~V.}}, \bibinfo {author} {\bibnamefont {Richardson},
  \bibfnamefont {J.~D.}}, \ and\ \bibinfo {author} {\bibnamefont {Tordella},
  \bibfnamefont {D.}},\ }\bibfield  {title} {\enquote {\bibinfo {title}
  {Magnetic turbulence spectra and intermittency in the heliosheath and in the
  local interstellar medium},}\ }\href {\doibase 10.3847/1538-4357/aafd30}
  {\bibfield  {journal} {\bibinfo  {journal} {The Astrophysical Journal}\
  }\textbf {\bibinfo {volume} {872}},\ \bibinfo {pages} {40} (\bibinfo {year}
  {2019})}\BibitemShut {NoStop}%
\bibitem [{\citenamefont {Frisch}(1995)}]{F1995}%
  \BibitemOpen
  \bibfield  {author} {\bibinfo {author} {\bibnamefont {Frisch}, \bibfnamefont
  {U.}},\ }\href@noop {} {\emph {\bibinfo {title} {Turbulence: The Legacy of A.
  N. Kolmogorov}}}\ (\bibinfo  {publisher} {Cambridge University Press.},\
  \bibinfo {year} {1995})\BibitemShut {NoStop}%
\bibitem [{\citenamefont {Galtier}\ and\ \citenamefont
  {Banerjee}(2011)}]{Ga2011}%
  \BibitemOpen
  \bibfield  {author} {\bibinfo {author} {\bibnamefont {Galtier}, \bibfnamefont
  {S.}}\ and\ \bibinfo {author} {\bibnamefont {Banerjee}, \bibfnamefont {S.}},\
  }\bibfield  {title} {\enquote {\bibinfo {title} {Exact relation for
  correlation functions in compressible isothermal turbulence},}\ }\href@noop
  {} {\bibfield  {journal} {\bibinfo  {journal} {Physical review letters}\
  }\textbf {\bibinfo {volume} {107}},\ \bibinfo {pages} {134501} (\bibinfo
  {year} {2011})}\BibitemShut {NoStop}%
\bibitem [{\citenamefont {Hadid}, \citenamefont {Sahraoui},\ and\ \citenamefont
  {Galtier}(2017)}]{H2017a}%
  \BibitemOpen
  \bibfield  {author} {\bibinfo {author} {\bibnamefont {Hadid}, \bibfnamefont
  {L.}}, \bibinfo {author} {\bibnamefont {Sahraoui}, \bibfnamefont {F.}}, \
  and\ \bibinfo {author} {\bibnamefont {Galtier}, \bibfnamefont {S.}},\
  }\bibfield  {title} {\enquote {\bibinfo {title} {Energy cascade rate in
  compressible fast and slow solar wind turbulence},}\ }\href@noop {}
  {\bibfield  {journal} {\bibinfo  {journal} {The Astrophysical Journal}\
  }\textbf {\bibinfo {volume} {838}},\ \bibinfo {pages} {9} (\bibinfo {year}
  {2017})}\BibitemShut {NoStop}%
\bibitem [{\citenamefont {Hadid}\ \emph {et~al.}(2018)\citenamefont {Hadid},
  \citenamefont {Sahraoui}, \citenamefont {Galtier},\ and\ \citenamefont
  {Huang}}]{Ha2017b}%
  \BibitemOpen
  \bibfield  {author} {\bibinfo {author} {\bibnamefont {Hadid}, \bibfnamefont
  {L.}}, \bibinfo {author} {\bibnamefont {Sahraoui}, \bibfnamefont {F.}},
  \bibinfo {author} {\bibnamefont {Galtier}, \bibfnamefont {S.}}, \ and\
  \bibinfo {author} {\bibnamefont {Huang}, \bibfnamefont {S.}},\ }\bibfield
  {title} {\enquote {\bibinfo {title} {Compressible magnetohydrodynamic
  turbulence in the earth's magnetosheath: estimation of the energy cascade
  rate using in situ spacecraft data},}\ }\href@noop {} {\bibfield  {journal}
  {\bibinfo  {journal} {Phys. Rev. Lett.}\ }\textbf {\bibinfo {volume} {120}},\
  \bibinfo {pages} {055102} (\bibinfo {year} {2018})}\BibitemShut {NoStop}%
\bibitem [{\citenamefont {Hernández}\ \emph {et~al.}(2021)\citenamefont
  {Hernández}, \citenamefont {Sorriso-Valvo}, \citenamefont {Bandyopadhyay},
  \citenamefont {Chasapis}, \citenamefont {Vásconez}, \citenamefont {Marino},\
  and\ \citenamefont {Pezzi}}]{He2021}%
  \BibitemOpen
  \bibfield  {author} {\bibinfo {author} {\bibnamefont {Hernández},
  \bibfnamefont {C.~S.}}, \bibinfo {author} {\bibnamefont {Sorriso-Valvo},
  \bibfnamefont {L.}}, \bibinfo {author} {\bibnamefont {Bandyopadhyay},
  \bibfnamefont {R.}}, \bibinfo {author} {\bibnamefont {Chasapis},
  \bibfnamefont {A.}}, \bibinfo {author} {\bibnamefont {Vásconez},
  \bibfnamefont {C.~L.}}, \bibinfo {author} {\bibnamefont {Marino},
  \bibfnamefont {R.}}, \ and\ \bibinfo {author} {\bibnamefont {Pezzi},
  \bibfnamefont {O.}},\ }\bibfield  {title} {\enquote {\bibinfo {title} {Impact
  of switchbacks on turbulent cascade and energy transfer rate in the inner
  heliosphere},}\ }\href {\doibase 10.3847/2041-8213/ac36d1} {\bibfield
  {journal} {\bibinfo  {journal} {The Astrophysical Journal Letters}\ }\textbf
  {\bibinfo {volume} {922}},\ \bibinfo {pages} {L11} (\bibinfo {year}
  {2021})}\BibitemShut {NoStop}%
\bibitem [{\citenamefont {Howes}\ \emph {et~al.}(2012)\citenamefont {Howes},
  \citenamefont {Bale}, \citenamefont {Klein}, \citenamefont {Chen},
  \citenamefont {Salem},\ and\ \citenamefont {TenBarge}}]{H2012}%
  \BibitemOpen
  \bibfield  {author} {\bibinfo {author} {\bibnamefont {Howes}, \bibfnamefont
  {G.~G.}}, \bibinfo {author} {\bibnamefont {Bale}, \bibfnamefont {S.~D.}},
  \bibinfo {author} {\bibnamefont {Klein}, \bibfnamefont {K.~G.}}, \bibinfo
  {author} {\bibnamefont {Chen}, \bibfnamefont {C.~H.~K.}}, \bibinfo {author}
  {\bibnamefont {Salem}, \bibfnamefont {C.~S.}}, \ and\ \bibinfo {author}
  {\bibnamefont {TenBarge}, \bibfnamefont {J.~M.}},\ }\bibfield  {title}
  {\enquote {\bibinfo {title} {The slow-mode nature of compressible wave power
  in solar wind turbulence},}\ }\href {\doibase 10.1088/2041-8205/753/1/l19}
  {\bibfield  {journal} {\bibinfo  {journal} {The Astrophysical Journal}\
  }\textbf {\bibinfo {volume} {753}},\ \bibinfo {pages} {L19} (\bibinfo {year}
  {2012})}\BibitemShut {NoStop}%
\bibitem [{\citenamefont {Huang}\ and\ \citenamefont {Sahraoui}(2019)}]{H2019}%
  \BibitemOpen
  \bibfield  {author} {\bibinfo {author} {\bibnamefont {Huang}, \bibfnamefont
  {S.}}\ and\ \bibinfo {author} {\bibnamefont {Sahraoui}, \bibfnamefont {F.}},\
  }\bibfield  {title} {\enquote {\bibinfo {title} {Testing of the taylor
  frozen-in-flow hypothesis at electron scales in the solar wind turbulence},}\
  }\href@noop {} {\bibfield  {journal} {\bibinfo  {journal} {The Astrophysical
  Journal}\ }\textbf {\bibinfo {volume} {876}},\ \bibinfo {pages} {138}
  (\bibinfo {year} {2019})}\BibitemShut {NoStop}%
\bibitem [{\citenamefont {Huang}\ \emph {et~al.}(2020)\citenamefont {Huang},
  \citenamefont {Zhang}, \citenamefont {Sahraoui}, \citenamefont {He},
  \citenamefont {Yuan}, \citenamefont {Andr{\'e}s}, \citenamefont {Hadid},
  \citenamefont {Deng}, \citenamefont {Jiang}, \citenamefont {Yu} \emph
  {et~al.}}]{H2020a}%
  \BibitemOpen
  \bibfield  {author} {\bibinfo {author} {\bibnamefont {Huang}, \bibfnamefont
  {S.}}, \bibinfo {author} {\bibnamefont {Zhang}, \bibfnamefont {J.}}, \bibinfo
  {author} {\bibnamefont {Sahraoui}, \bibfnamefont {F.}}, \bibinfo {author}
  {\bibnamefont {He}, \bibfnamefont {J.}}, \bibinfo {author} {\bibnamefont
  {Yuan}, \bibfnamefont {Z.}}, \bibinfo {author} {\bibnamefont {Andr{\'e}s},
  \bibfnamefont {N.}}, \bibinfo {author} {\bibnamefont {Hadid}, \bibfnamefont
  {L.}}, \bibinfo {author} {\bibnamefont {Deng}, \bibfnamefont {X.}}, \bibinfo
  {author} {\bibnamefont {Jiang}, \bibfnamefont {K.}}, \bibinfo {author}
  {\bibnamefont {Yu}, \bibfnamefont {L.}},  \emph {et~al.},\ }\bibfield
  {title} {\enquote {\bibinfo {title} {Kinetic scale slow solar wind turbulence
  in the inner heliosphere: Coexistence of kinetic alfv{\'e}n waves and
  alfv{\'e}n ion cyclotron waves},}\ }\href@noop {} {\bibfield  {journal}
  {\bibinfo  {journal} {The Astrophysical Journal Letters}\ }\textbf {\bibinfo
  {volume} {897}},\ \bibinfo {pages} {L3} (\bibinfo {year} {2020})}\BibitemShut
  {NoStop}%
\bibitem [{\citenamefont {von K\'arm\'an}\ and\ \citenamefont
  {Howarth}(1938)}]{vkh1938}%
  \BibitemOpen
  \bibfield  {author} {\bibinfo {author} {\bibnamefont {von K\'arm\'an},
  \bibfnamefont {T.}}\ and\ \bibinfo {author} {\bibnamefont {Howarth},
  \bibfnamefont {L.}},\ }\bibfield  {title} {\enquote {\bibinfo {title} {On the
  statistical theory of isotropic turbulence},}\ }\href {\doibase
  10.1098/rspa.1938.0013} {\bibfield  {journal} {\bibinfo  {journal}
  {Proceedings of the Royal Society of London A: Mathematical, Physical and
  Engineering Sciences}\ }\textbf {\bibinfo {volume} {164}},\ \bibinfo {pages}
  {192--215} (\bibinfo {year} {1938})}\BibitemShut {NoStop}%
\bibitem [{\citenamefont {Kasper}\ \emph {et~al.}(2016)\citenamefont {Kasper},
  \citenamefont {Abiad}, \citenamefont {Austin}, \citenamefont
  {Balat-Pichelin}, \citenamefont {Bale}, \citenamefont {Belcher},
  \citenamefont {Berg}, \citenamefont {Bergner}, \citenamefont {Berthomier},
  \citenamefont {Bookbinder} \emph {et~al.}}]{K2016}%
  \BibitemOpen
  \bibfield  {author} {\bibinfo {author} {\bibnamefont {Kasper}, \bibfnamefont
  {J.~C.}}, \bibinfo {author} {\bibnamefont {Abiad}, \bibfnamefont {R.}},
  \bibinfo {author} {\bibnamefont {Austin}, \bibfnamefont {G.}}, \bibinfo
  {author} {\bibnamefont {Balat-Pichelin}, \bibfnamefont {M.}}, \bibinfo
  {author} {\bibnamefont {Bale}, \bibfnamefont {S.~D.}}, \bibinfo {author}
  {\bibnamefont {Belcher}, \bibfnamefont {J.~W.}}, \bibinfo {author}
  {\bibnamefont {Berg}, \bibfnamefont {P.}}, \bibinfo {author} {\bibnamefont
  {Bergner}, \bibfnamefont {H.}}, \bibinfo {author} {\bibnamefont {Berthomier},
  \bibfnamefont {M.}}, \bibinfo {author} {\bibnamefont {Bookbinder},
  \bibfnamefont {J.}},  \emph {et~al.},\ }\bibfield  {title} {\enquote
  {\bibinfo {title} {Solar wind electrons alphas and protons (sweap)
  investigation: design of the solar wind and coronal plasma instrument suite
  for solar probe plus},}\ }\href@noop {} {\bibfield  {journal} {\bibinfo
  {journal} {Space Science Reviews}\ }\textbf {\bibinfo {volume} {204}},\
  \bibinfo {pages} {131--186} (\bibinfo {year} {2016})}\BibitemShut {NoStop}%
\bibitem [{\citenamefont {Kasper}\ \emph {et~al.}(2019)\citenamefont {Kasper},
  \citenamefont {Bale}, \citenamefont {Belcher}, \citenamefont {Berthomier},
  \citenamefont {Case}, \citenamefont {Chandran}, \citenamefont {Curtis},
  \citenamefont {Gallagher}, \citenamefont {Gary}, \citenamefont {Golub} \emph
  {et~al.}}]{Ka2019}%
  \BibitemOpen
  \bibfield  {author} {\bibinfo {author} {\bibnamefont {Kasper}, \bibfnamefont
  {J.~C.}}, \bibinfo {author} {\bibnamefont {Bale}, \bibfnamefont {S.~D.}},
  \bibinfo {author} {\bibnamefont {Belcher}, \bibfnamefont {J.~W.}}, \bibinfo
  {author} {\bibnamefont {Berthomier}, \bibfnamefont {M.}}, \bibinfo {author}
  {\bibnamefont {Case}, \bibfnamefont {A.~W.}}, \bibinfo {author} {\bibnamefont
  {Chandran}, \bibfnamefont {B.~D.}}, \bibinfo {author} {\bibnamefont {Curtis},
  \bibfnamefont {D.}}, \bibinfo {author} {\bibnamefont {Gallagher},
  \bibfnamefont {D.}}, \bibinfo {author} {\bibnamefont {Gary}, \bibfnamefont
  {S.}}, \bibinfo {author} {\bibnamefont {Golub}, \bibfnamefont {L.}},  \emph
  {et~al.},\ }\bibfield  {title} {\enquote {\bibinfo {title} {Alfv{\'e}nic
  velocity spikes and rotational flows in the near-sun solar wind},}\
  }\href@noop {} {\bibfield  {journal} {\bibinfo  {journal} {Nature}\ }\textbf
  {\bibinfo {volume} {576}},\ \bibinfo {pages} {228--231} (\bibinfo {year}
  {2019})}\BibitemShut {NoStop}%
\bibitem [{\citenamefont {Kiyani}, \citenamefont {Osman},\ and\ \citenamefont
  {Chapman}(2015)}]{K2015}%
  \BibitemOpen
  \bibfield  {author} {\bibinfo {author} {\bibnamefont {Kiyani}, \bibfnamefont
  {K.~H.}}, \bibinfo {author} {\bibnamefont {Osman}, \bibfnamefont {K.~T.}}, \
  and\ \bibinfo {author} {\bibnamefont {Chapman}, \bibfnamefont {S.~C.}},\
  }\href@noop {} {\enquote {\bibinfo {title} {Dissipation and heating in solar
  wind turbulence: from the macro to the micro and back again},}\ } (\bibinfo
  {year} {2015})\BibitemShut {NoStop}%
\bibitem [{\citenamefont {Kritsuk}, \citenamefont {Wagner},\ and\ \citenamefont
  {Norman}(2013)}]{Kr2013}%
  \BibitemOpen
  \bibfield  {author} {\bibinfo {author} {\bibnamefont {Kritsuk}, \bibfnamefont
  {A.~G.}}, \bibinfo {author} {\bibnamefont {Wagner}, \bibfnamefont {R.}}, \
  and\ \bibinfo {author} {\bibnamefont {Norman}, \bibfnamefont {M.~L.}},\
  }\bibfield  {title} {\enquote {\bibinfo {title} {Energy cascade and scaling
  in supersonic isothermal turbulence},}\ }\href@noop {} {\bibfield  {journal}
  {\bibinfo  {journal} {Journal of Fluid Mechanics}\ }\textbf {\bibinfo
  {volume} {729}},\ \bibinfo {pages} {R1} (\bibinfo {year} {2013})}\BibitemShut
  {NoStop}%
\bibitem [{\citenamefont {Liu}, \citenamefont {Shah},\ and\ \citenamefont
  {Jiang}(2004)}]{LIU2004}%
  \BibitemOpen
  \bibfield  {author} {\bibinfo {author} {\bibnamefont {Liu}, \bibfnamefont
  {H.}}, \bibinfo {author} {\bibnamefont {Shah}, \bibfnamefont {S.}}, \ and\
  \bibinfo {author} {\bibnamefont {Jiang}, \bibfnamefont {W.}},\ }\bibfield
  {title} {\enquote {\bibinfo {title} {On-line outlier detection and data
  cleaning},}\ }\href {\doibase
  https://doi.org/10.1016/j.compchemeng.2004.01.009} {\bibfield  {journal}
  {\bibinfo  {journal} {Computers and Chemical Engineering}\ }\textbf {\bibinfo
  {volume} {28}},\ \bibinfo {pages} {1635--1647} (\bibinfo {year}
  {2004})}\BibitemShut {NoStop}%
\bibitem [{\citenamefont {{MacBride}}, \citenamefont {{Forman}},\ and\
  \citenamefont {{Smith}}(2005)}]{MacBride2005}%
  \BibitemOpen
  \bibfield  {author} {\bibinfo {author} {\bibnamefont {{MacBride}},
  \bibfnamefont {B.~T.}}, \bibinfo {author} {\bibnamefont {{Forman}},
  \bibfnamefont {M.~A.}}, \ and\ \bibinfo {author} {\bibnamefont {{Smith}},
  \bibfnamefont {C.~W.}},\ }\bibfield  {title} {\enquote {\bibinfo {title}
  {{Turbulence and Third Moment of Fluctuations: Kolmogorov's 4/5 Law and its
  MHD Analogues in the Solar Wind}},}\ }\bibfield  {booktitle} {\emph {\bibinfo
  {booktitle} {Solar Wind 11/SOHO 16, Connecting Sun and Heliosphere}},\
  }\href@noop {} {\ \bibinfo {series} {ESA Special Publication},\ \textbf
  {\bibinfo {volume} {592}},\ \bibinfo {pages} {613} (\bibinfo {year}
  {2005})}\BibitemShut {NoStop}%
\bibitem [{\citenamefont {MacBride}, \citenamefont {Smith},\ and\ \citenamefont
  {Forman}(2008)}]{Mc2008}%
  \BibitemOpen
  \bibfield  {author} {\bibinfo {author} {\bibnamefont {MacBride},
  \bibfnamefont {B.~T.}}, \bibinfo {author} {\bibnamefont {Smith},
  \bibfnamefont {C.~W.}}, \ and\ \bibinfo {author} {\bibnamefont {Forman},
  \bibfnamefont {M.~A.}},\ }\href@noop {} {\bibfield  {journal} {\bibinfo
  {journal} {Astrophys. J.}\ }\textbf {\bibinfo {volume} {679}},\ \bibinfo
  {pages} {1644--1660} (\bibinfo {year} {2008})}\BibitemShut {NoStop}%
\bibitem [{\citenamefont {Marino}\ and\ \citenamefont
  {Sorriso-Valvo}(2023)}]{Ma2023}%
  \BibitemOpen
  \bibfield  {author} {\bibinfo {author} {\bibnamefont {Marino}, \bibfnamefont
  {R.}}\ and\ \bibinfo {author} {\bibnamefont {Sorriso-Valvo}, \bibfnamefont
  {L.}},\ }\bibfield  {title} {\enquote {\bibinfo {title} {Scaling laws for the
  energy transfer in space plasma turbulence},}\ }\href@noop {} {\bibfield
  {journal} {\bibinfo  {journal} {Physics Reports}\ }\textbf {\bibinfo {volume}
  {1006}},\ \bibinfo {pages} {1--144} (\bibinfo {year} {2023})}\BibitemShut
  {NoStop}%
\bibitem [{\citenamefont {Marino}\ \emph {et~al.}(2008)\citenamefont {Marino},
  \citenamefont {Sorriso-Valvo}, \citenamefont {Carbone}, \citenamefont
  {Noullez}, \citenamefont {Bruno},\ and\ \citenamefont {Bavassano}}]{M2008}%
  \BibitemOpen
  \bibfield  {author} {\bibinfo {author} {\bibnamefont {Marino}, \bibfnamefont
  {R.}}, \bibinfo {author} {\bibnamefont {Sorriso-Valvo}, \bibfnamefont {L.}},
  \bibinfo {author} {\bibnamefont {Carbone}, \bibfnamefont {V.}}, \bibinfo
  {author} {\bibnamefont {Noullez}, \bibfnamefont {A.}}, \bibinfo {author}
  {\bibnamefont {Bruno}, \bibfnamefont {R.}}, \ and\ \bibinfo {author}
  {\bibnamefont {Bavassano}, \bibfnamefont {B.}},\ }\bibfield  {title}
  {\enquote {\bibinfo {title} {{Heating the Solar Wind by a Magnetohydrodynamic
  Turbulent Energy Cascade}},}\ }\href@noop {} {\bibfield  {journal} {\bibinfo
  {journal} {Astrophys. J. Lett.}\ }\textbf {\bibinfo {volume} {677}},\
  \bibinfo {pages} {L71--L74} (\bibinfo {year} {2008})}\BibitemShut {NoStop}%
\bibitem [{\citenamefont {Marsch}\ and\ \citenamefont
  {Mangeney}(1987)}]{M1987}%
  \BibitemOpen
  \bibfield  {author} {\bibinfo {author} {\bibnamefont {Marsch}, \bibfnamefont
  {E.}}\ and\ \bibinfo {author} {\bibnamefont {Mangeney}, \bibfnamefont {A.}},\
  }\bibfield  {title} {\enquote {\bibinfo {title} {Ideal mhd equations in terms
  of compressive elsässer variables},}\ }\href {\doibase
  10.1029/JA092iA07p07363} {\bibfield  {journal} {\bibinfo  {journal} {Journal
  of Geophysical Research: Space Physics}\ }\textbf {\bibinfo {volume} {92}},\
  \bibinfo {pages} {7363--7367} (\bibinfo {year} {1987})}\BibitemShut {NoStop}%
\bibitem [{\citenamefont {Marsch}\ \emph {et~al.}(1982)\citenamefont {Marsch},
  \citenamefont {Mühlhäuser}, \citenamefont {Schwenn}, \citenamefont
  {Rosenbauer}, \citenamefont {Pilipp},\ and\ \citenamefont
  {Neubauer}}]{M1982c}%
  \BibitemOpen
  \bibfield  {author} {\bibinfo {author} {\bibnamefont {Marsch}, \bibfnamefont
  {E.}}, \bibinfo {author} {\bibnamefont {Mühlhäuser}, \bibfnamefont
  {K.-H.}}, \bibinfo {author} {\bibnamefont {Schwenn}, \bibfnamefont {R.}},
  \bibinfo {author} {\bibnamefont {Rosenbauer}, \bibfnamefont {H.}}, \bibinfo
  {author} {\bibnamefont {Pilipp}, \bibfnamefont {W.}}, \ and\ \bibinfo
  {author} {\bibnamefont {Neubauer}, \bibfnamefont {F.~M.}},\ }\bibfield
  {title} {\enquote {\bibinfo {title} {Solar wind protons: Three-dimensional
  velocity distributions and derived plasma parameters measured between 0.3 and
  1 au},}\ }\href {\doibase 10.1029/JA087iA01p00052} {\bibfield  {journal}
  {\bibinfo  {journal} {Journal of Geophysical Research: Space Physics}\
  }\textbf {\bibinfo {volume} {87}},\ \bibinfo {pages} {52--72} (\bibinfo
  {year} {1982})},\ \Eprint
  {http://arxiv.org/abs/https://agupubs.onlinelibrary.wiley.com/doi/pdf/10.1029/JA087iA01p00052}
  {https://agupubs.onlinelibrary.wiley.com/doi/pdf/10.1029/JA087iA01p00052}
  \BibitemShut {NoStop}%
\bibitem [{\citenamefont {{Masters}}\ \emph {et~al.}(2008)\citenamefont
  {{Masters}}, \citenamefont {{Achilleos}}, \citenamefont {{Dougherty}},
  \citenamefont {{Slavin}}, \citenamefont {{Hospodarsky}}, \citenamefont
  {{Arridge}},\ and\ \citenamefont {{Coates}}}]{Ma2008}%
  \BibitemOpen
  \bibfield  {author} {\bibinfo {author} {\bibnamefont {{Masters}},
  \bibfnamefont {A.}}, \bibinfo {author} {\bibnamefont {{Achilleos}},
  \bibfnamefont {N.}}, \bibinfo {author} {\bibnamefont {{Dougherty}},
  \bibfnamefont {M.~K.}}, \bibinfo {author} {\bibnamefont {{Slavin}},
  \bibfnamefont {J.~A.}}, \bibinfo {author} {\bibnamefont {{Hospodarsky}},
  \bibfnamefont {G.~B.}}, \bibinfo {author} {\bibnamefont {{Arridge}},
  \bibfnamefont {C.~S.}}, \ and\ \bibinfo {author} {\bibnamefont {{Coates}},
  \bibfnamefont {A.~J.}},\ }\bibfield  {title} {\enquote {\bibinfo {title} {{An
  empirical model of Saturn's bow shock: Cassini observations of shock location
  and shape}},}\ }\href {\doibase 10.1029/2008JA013276} {\bibfield  {journal}
  {\bibinfo  {journal} {Journal of Geophysical Research (Space Physics)}\
  }\textbf {\bibinfo {volume} {113}},\ \bibinfo {eid} {A10210} (\bibinfo {year}
  {2008})}\BibitemShut {NoStop}%
\bibitem [{\citenamefont {Matthaeus}\ and\ \citenamefont
  {Velli}(2011)}]{M2011}%
  \BibitemOpen
  \bibfield  {author} {\bibinfo {author} {\bibnamefont {Matthaeus},
  \bibfnamefont {W.}}\ and\ \bibinfo {author} {\bibnamefont {Velli},
  \bibfnamefont {M.}},\ }\bibfield  {title} {\enquote {\bibinfo {title} {Who
  needs turbulence?}}\ }\href@noop {} {\bibfield  {journal} {\bibinfo
  {journal} {Space science reviews}\ }\textbf {\bibinfo {volume} {160}},\
  \bibinfo {pages} {145} (\bibinfo {year} {2011})}\BibitemShut {NoStop}%
\bibitem [{\citenamefont {Matthaeus}\ \emph {et~al.}(1999)\citenamefont
  {Matthaeus}, \citenamefont {Zank}, \citenamefont {Smith},\ and\ \citenamefont
  {Oughton}}]{M1999}%
  \BibitemOpen
  \bibfield  {author} {\bibinfo {author} {\bibnamefont {Matthaeus},
  \bibfnamefont {W.~H.}}, \bibinfo {author} {\bibnamefont {Zank}, \bibfnamefont
  {G.~P.}}, \bibinfo {author} {\bibnamefont {Smith}, \bibfnamefont {C.~W.}}, \
  and\ \bibinfo {author} {\bibnamefont {Oughton}, \bibfnamefont {S.}},\
  }\href@noop {} {\bibfield  {journal} {\bibinfo  {journal} {Phys. Rev. Lett.}\
  }\textbf {\bibinfo {volume} {82}},\ \bibinfo {pages} {3444--} (\bibinfo
  {year} {1999})}\BibitemShut {NoStop}%
\bibitem [{\citenamefont {Mininni}\ and\ \citenamefont
  {Pouquet}(2009)}]{Mi2009}%
  \BibitemOpen
  \bibfield  {author} {\bibinfo {author} {\bibnamefont {Mininni}, \bibfnamefont
  {P.~D.}}\ and\ \bibinfo {author} {\bibnamefont {Pouquet}, \bibfnamefont
  {A.}},\ }\bibfield  {title} {\enquote {\bibinfo {title} {Finite dissipation
  and intermittency in magnetohydrodynamics},}\ }\href@noop {} {\bibfield
  {journal} {\bibinfo  {journal} {Phys. Rev. E}\ }\textbf {\bibinfo {volume}
  {80}},\ \bibinfo {pages} {025401} (\bibinfo {year} {2009})}\BibitemShut
  {NoStop}%
\bibitem [{\citenamefont {Monin}\ and\ \citenamefont {Yaglom}(1975)}]{MY1975}%
  \BibitemOpen
  \bibfield  {author} {\bibinfo {author} {\bibnamefont {Monin}, \bibfnamefont
  {A.~S.}}\ and\ \bibinfo {author} {\bibnamefont {Yaglom}, \bibfnamefont
  {A.~M.}},\ }\href@noop {} {\emph {\bibinfo {title} {Statistical Fluid
  Mechanics: Mechanics of Turbulence}}},\ Vol.~\bibinfo {volume} {2}\ (\bibinfo
   {publisher} {Cambridge, MA: MIT Press.},\ \bibinfo {year}
  {1975})\BibitemShut {NoStop}%
\bibitem [{\citenamefont {Osman}\ \emph {et~al.}(2011)\citenamefont {Osman},
  \citenamefont {Wan}, \citenamefont {Matthaeus}, \citenamefont {Weygand},\
  and\ \citenamefont {Dasso}}]{O2011}%
  \BibitemOpen
  \bibfield  {author} {\bibinfo {author} {\bibnamefont {Osman}, \bibfnamefont
  {K.~T.}}, \bibinfo {author} {\bibnamefont {Wan}, \bibfnamefont {M.}},
  \bibinfo {author} {\bibnamefont {Matthaeus}, \bibfnamefont {W.~H.}}, \bibinfo
  {author} {\bibnamefont {Weygand}, \bibfnamefont {J.~M.}}, \ and\ \bibinfo
  {author} {\bibnamefont {Dasso}, \bibfnamefont {S.}},\ }\bibfield  {title}
  {\enquote {\bibinfo {title} {Anisotropic third-moment estimates of the energy
  cascade in solar wind turbulence using multispacecraft data},}\ }\href
  {\doibase 10.1103/PhysRevLett.107.165001} {\bibfield  {journal} {\bibinfo
  {journal} {Phys. Rev. Lett.}\ }\textbf {\bibinfo {volume} {107}},\ \bibinfo
  {pages} {165001} (\bibinfo {year} {2011})}\BibitemShut {NoStop}%
\bibitem [{\citenamefont {Parashar}\ \emph {et~al.}(2020)\citenamefont
  {Parashar}, \citenamefont {Goldstein}, \citenamefont {Maruca}, \citenamefont
  {Matthaeus}, \citenamefont {Ruffolo}, \citenamefont {Bandyopadhyay},
  \citenamefont {Chhiber}, \citenamefont {Chasapis}, \citenamefont {Qudsi},
  \citenamefont {Vech} \emph {et~al.}}]{P2020}%
  \BibitemOpen
  \bibfield  {author} {\bibinfo {author} {\bibnamefont {Parashar},
  \bibfnamefont {T.}}, \bibinfo {author} {\bibnamefont {Goldstein},
  \bibfnamefont {M.}}, \bibinfo {author} {\bibnamefont {Maruca}, \bibfnamefont
  {B.}}, \bibinfo {author} {\bibnamefont {Matthaeus}, \bibfnamefont {W.}},
  \bibinfo {author} {\bibnamefont {Ruffolo}, \bibfnamefont {D.}}, \bibinfo
  {author} {\bibnamefont {Bandyopadhyay}, \bibfnamefont {R.}}, \bibinfo
  {author} {\bibnamefont {Chhiber}, \bibfnamefont {R.}}, \bibinfo {author}
  {\bibnamefont {Chasapis}, \bibfnamefont {A.}}, \bibinfo {author}
  {\bibnamefont {Qudsi}, \bibfnamefont {R.}}, \bibinfo {author} {\bibnamefont
  {Vech}, \bibfnamefont {D.}},  \emph {et~al.},\ }\bibfield  {title} {\enquote
  {\bibinfo {title} {Measures of scale-dependent alfv{\'e}nicity in the first
  psp solar encounter},}\ }\href@noop {} {\bibfield  {journal} {\bibinfo
  {journal} {The Astrophysical Journal Supplement Series}\ }\textbf {\bibinfo
  {volume} {246}},\ \bibinfo {pages} {58} (\bibinfo {year} {2020})}\BibitemShut
  {NoStop}%
\bibitem [{\citenamefont {Pearson}\ \emph {et~al.}(2016)\citenamefont
  {Pearson}, \citenamefont {Neuvo}, \citenamefont {Astola},\ and\ \citenamefont
  {Gabbouj}}]{Pearson2016}%
  \BibitemOpen
  \bibfield  {author} {\bibinfo {author} {\bibnamefont {Pearson}, \bibfnamefont
  {R.}}, \bibinfo {author} {\bibnamefont {Neuvo}, \bibfnamefont {Y.}}, \bibinfo
  {author} {\bibnamefont {Astola}, \bibfnamefont {J.}}, \ and\ \bibinfo
  {author} {\bibnamefont {Gabbouj}, \bibfnamefont {M.}},\ }\bibfield  {title}
  {\enquote {\bibinfo {title} {Generalized hampel filters},}\ }\href {\doibase
  10.1186/s13634-016-0383-6} {\bibfield  {journal} {\bibinfo  {journal}
  {EURASIP Journal on Advances in Signal Processing}\ }\textbf {\bibinfo
  {volume} {2016}} (\bibinfo {year} {2016}),\
  10.1186/s13634-016-0383-6}\BibitemShut {NoStop}%
\bibitem [{\citenamefont {Pine}\ \emph {et~al.}(2020)\citenamefont {Pine},
  \citenamefont {Smith}, \citenamefont {Hollick}, \citenamefont {Argall},
  \citenamefont {Vasquez}, \citenamefont {Isenberg}, \citenamefont {Schwadron},
  \citenamefont {Joyce}, \citenamefont {Sok{\'{o}}{\l}}, \citenamefont
  {Bzowski}, \citenamefont {Kubiak}, \citenamefont {Hamilton}, \citenamefont
  {McLaurin},\ and\ \citenamefont {Leamon}}]{Pi2020}%
  \BibitemOpen
  \bibfield  {author} {\bibinfo {author} {\bibnamefont {Pine}, \bibfnamefont
  {Z.~B.}}, \bibinfo {author} {\bibnamefont {Smith}, \bibfnamefont {C.~W.}},
  \bibinfo {author} {\bibnamefont {Hollick}, \bibfnamefont {S.~J.}}, \bibinfo
  {author} {\bibnamefont {Argall}, \bibfnamefont {M.~R.}}, \bibinfo {author}
  {\bibnamefont {Vasquez}, \bibfnamefont {B.~J.}}, \bibinfo {author}
  {\bibnamefont {Isenberg}, \bibfnamefont {P.~A.}}, \bibinfo {author}
  {\bibnamefont {Schwadron}, \bibfnamefont {N.~A.}}, \bibinfo {author}
  {\bibnamefont {Joyce}, \bibfnamefont {C.~J.}}, \bibinfo {author}
  {\bibnamefont {Sok{\'{o}}{\l}}, \bibfnamefont {J.~M.}}, \bibinfo {author}
  {\bibnamefont {Bzowski}, \bibfnamefont {M.}}, \bibinfo {author} {\bibnamefont
  {Kubiak}, \bibfnamefont {M.~A.}}, \bibinfo {author} {\bibnamefont {Hamilton},
  \bibfnamefont {K.~E.}}, \bibinfo {author} {\bibnamefont {McLaurin},
  \bibfnamefont {M.~L.}}, \ and\ \bibinfo {author} {\bibnamefont {Leamon},
  \bibfnamefont {R.~J.}},\ }\bibfield  {title} {\enquote {\bibinfo {title}
  {Solar wind turbulence from 1 to 45 au. i. evidence for dissipation of
  magnetic fluctuations using voyager and {ACE} observations},}\ }\href
  {\doibase 10.3847/1538-4357/abab10} {\bibfield  {journal} {\bibinfo
  {journal} {The Astrophysical Journal}\ }\textbf {\bibinfo {volume} {900}},\
  \bibinfo {pages} {91} (\bibinfo {year} {2020})}\BibitemShut {NoStop}%
\bibitem [{\citenamefont {Politano}\ and\ \citenamefont
  {Pouquet}(1998{\natexlab{a}})}]{P1998b}%
  \BibitemOpen
  \bibfield  {author} {\bibinfo {author} {\bibnamefont {Politano},
  \bibfnamefont {H.}}\ and\ \bibinfo {author} {\bibnamefont {Pouquet},
  \bibfnamefont {A.}},\ }\bibfield  {title} {\enquote {\bibinfo {title}
  {Dynamical length scales for turbulent magnetized flows},}\ }\href@noop {}
  {\bibfield  {journal} {\bibinfo  {journal} {Geophysical Research Letters}\
  }\textbf {\bibinfo {volume} {25}},\ \bibinfo {pages} {273--276} (\bibinfo
  {year} {1998}{\natexlab{a}})}\BibitemShut {NoStop}%
\bibitem [{\citenamefont {Politano}\ and\ \citenamefont
  {Pouquet}(1998{\natexlab{b}})}]{P1998a}%
  \BibitemOpen
  \bibfield  {author} {\bibinfo {author} {\bibnamefont {Politano},
  \bibfnamefont {H.}}\ and\ \bibinfo {author} {\bibnamefont {Pouquet},
  \bibfnamefont {A.}},\ }\bibfield  {title} {\enquote {\bibinfo {title} {von
  k{\'a}rm{\'a}n--howarth equation for magnetohydrodynamics and its
  consequences on third-order longitudinal structure and correlation
  functions},}\ }\href@noop {} {\bibfield  {journal} {\bibinfo  {journal}
  {Physical Review E}\ }\textbf {\bibinfo {volume} {57}},\ \bibinfo {pages}
  {R21} (\bibinfo {year} {1998}{\natexlab{b}})}\BibitemShut {NoStop}%
\bibitem [{\citenamefont {Sahraoui}(2008)}]{Sa2008}%
  \BibitemOpen
  \bibfield  {author} {\bibinfo {author} {\bibnamefont {Sahraoui},
  \bibfnamefont {F.}},\ }\bibfield  {title} {\enquote {\bibinfo {title}
  {Diagnosis of magnetic structures and intermittency in space-plasma
  turbulence using the technique of surrogate data},}\ }\href@noop {}
  {\bibfield  {journal} {\bibinfo  {journal} {Phys. Rev. E}\ }\textbf {\bibinfo
  {volume} {78}},\ \bibinfo {pages} {026402} (\bibinfo {year}
  {2008})}\BibitemShut {NoStop}%
\bibitem [{\citenamefont {Sahraoui}, \citenamefont {Hadid},\ and\ \citenamefont
  {Huang}(2020)}]{S2020}%
  \BibitemOpen
  \bibfield  {author} {\bibinfo {author} {\bibnamefont {Sahraoui},
  \bibfnamefont {F.}}, \bibinfo {author} {\bibnamefont {Hadid}, \bibfnamefont
  {L.}}, \ and\ \bibinfo {author} {\bibnamefont {Huang}, \bibfnamefont {S.}},\
  }\bibfield  {title} {\enquote {\bibinfo {title} {Magnetohydrodynamic and
  kinetic scale turbulence in the near-earth space plasmas: a (short) biased
  review},}\ }\href@noop {} {\bibfield  {journal} {\bibinfo  {journal} {Reviews
  of Modern Plasma Physics}\ }\textbf {\bibinfo {volume} {4}},\ \bibinfo
  {pages} {1--33} (\bibinfo {year} {2020})}\BibitemShut {NoStop}%
\bibitem [{\citenamefont {Simakov}\ and\ \citenamefont
  {Chac\'on}(2008)}]{Si2008}%
  \BibitemOpen
  \bibfield  {author} {\bibinfo {author} {\bibnamefont {Simakov}, \bibfnamefont
  {A.~N.}}\ and\ \bibinfo {author} {\bibnamefont {Chac\'on}, \bibfnamefont
  {L.}},\ }\bibfield  {title} {\enquote {\bibinfo {title} {Quantitative,
  comprehensive, analytical model for magnetic reconnection in hall
  magnetohydrodynamics},}\ }\href {\doibase 10.1103/PhysRevLett.101.105003}
  {\bibfield  {journal} {\bibinfo  {journal} {Phys. Rev. Lett.}\ }\textbf
  {\bibinfo {volume} {101}},\ \bibinfo {pages} {105003} (\bibinfo {year}
  {2008})}\BibitemShut {NoStop}%
\bibitem [{\citenamefont {Simon}\ and\ \citenamefont {Sahraoui}(2021)}]{S2021}%
  \BibitemOpen
  \bibfield  {author} {\bibinfo {author} {\bibnamefont {Simon}, \bibfnamefont
  {P.}}\ and\ \bibinfo {author} {\bibnamefont {Sahraoui}, \bibfnamefont {F.}},\
  }\bibfield  {title} {\enquote {\bibinfo {title} {General exact law of
  compressible isentropic magnetohydrodynamic flows: theory and spacecraft
  observations in the solar wind},}\ }\href@noop {} {\bibfield  {journal}
  {\bibinfo  {journal} {The Astrophysical Journal}\ }\textbf {\bibinfo {volume}
  {916}},\ \bibinfo {pages} {49} (\bibinfo {year} {2021})}\BibitemShut
  {NoStop}%
\bibitem [{\citenamefont {Simon}\ and\ \citenamefont {Sahraoui}(2022)}]{S2022}%
  \BibitemOpen
  \bibfield  {author} {\bibinfo {author} {\bibnamefont {Simon}, \bibfnamefont
  {P.}}\ and\ \bibinfo {author} {\bibnamefont {Sahraoui}, \bibfnamefont {F.}},\
  }\bibfield  {title} {\enquote {\bibinfo {title} {Exact law for compressible
  pressure-anisotropic magnetohydrodynamic turbulence: Toward linking energy
  cascade and instabilities},}\ }\href {\doibase 10.1103/PhysRevE.105.055111}
  {\bibfield  {journal} {\bibinfo  {journal} {Phys. Rev. E}\ }\textbf {\bibinfo
  {volume} {105}},\ \bibinfo {pages} {055111} (\bibinfo {year}
  {2022})}\BibitemShut {NoStop}%
\bibitem [{\citenamefont {Sorriso-Valvo}\ \emph {et~al.}(2002)\citenamefont
  {Sorriso-Valvo}, \citenamefont {Carbone}, \citenamefont {Noullez},
  \citenamefont {Politano}, \citenamefont {Pouquet},\ and\ \citenamefont
  {Veltri}}]{Sorriso2002}%
  \BibitemOpen
  \bibfield  {author} {\bibinfo {author} {\bibnamefont {Sorriso-Valvo},
  \bibfnamefont {L.}}, \bibinfo {author} {\bibnamefont {Carbone}, \bibfnamefont
  {V.}}, \bibinfo {author} {\bibnamefont {Noullez}, \bibfnamefont {A.}},
  \bibinfo {author} {\bibnamefont {Politano}, \bibfnamefont {H.}}, \bibinfo
  {author} {\bibnamefont {Pouquet}, \bibfnamefont {A.}}, \ and\ \bibinfo
  {author} {\bibnamefont {Veltri}, \bibfnamefont {P.}},\ }\bibfield  {title}
  {\enquote {\bibinfo {title} {Analysis of cancellation in two-dimensional
  magnetohydrodynamic turbulence},}\ }\href {\doibase 10.1063/1.1420738}
  {\bibfield  {journal} {\bibinfo  {journal} {Physics of Plasmas}\ }\textbf
  {\bibinfo {volume} {9}},\ \bibinfo {pages} {89--95} (\bibinfo {year}
  {2002})}\BibitemShut {NoStop}%
\bibitem [{\citenamefont {Sorriso-Valvo}\ \emph {et~al.}(2007)\citenamefont
  {Sorriso-Valvo}, \citenamefont {Marino}, \citenamefont {Carbone},
  \citenamefont {Noullez}, \citenamefont {Lepreti}, \citenamefont {Veltri},
  \citenamefont {Bruno}, \citenamefont {Bavassano},\ and\ \citenamefont
  {Pietropaolo}}]{So2007}%
  \BibitemOpen
  \bibfield  {author} {\bibinfo {author} {\bibnamefont {Sorriso-Valvo},
  \bibfnamefont {L.}}, \bibinfo {author} {\bibnamefont {Marino}, \bibfnamefont
  {R.}}, \bibinfo {author} {\bibnamefont {Carbone}, \bibfnamefont {V.}},
  \bibinfo {author} {\bibnamefont {Noullez}, \bibfnamefont {A.}}, \bibinfo
  {author} {\bibnamefont {Lepreti}, \bibfnamefont {F.}}, \bibinfo {author}
  {\bibnamefont {Veltri}, \bibfnamefont {P.}}, \bibinfo {author} {\bibnamefont
  {Bruno}, \bibfnamefont {R.}}, \bibinfo {author} {\bibnamefont {Bavassano},
  \bibfnamefont {B.}}, \ and\ \bibinfo {author} {\bibnamefont {Pietropaolo},
  \bibfnamefont {E.}},\ }\bibfield  {title} {\enquote {\bibinfo {title}
  {Observation of inertial energy cascade in interplanetary space plasma},}\
  }\href {\doibase 10.1103/PhysRevLett.99.115001} {\bibfield  {journal}
  {\bibinfo  {journal} {Phys. Rev. Lett.}\ }\textbf {\bibinfo {volume} {99}},\
  \bibinfo {pages} {115001} (\bibinfo {year} {2007})}\BibitemShut {NoStop}%
\bibitem [{\citenamefont {Stawarz}\ \emph {et~al.}(2009)\citenamefont
  {Stawarz}, \citenamefont {Smith}, \citenamefont {Vasquez}, \citenamefont
  {Forman},\ and\ \citenamefont {MacBride}}]{St2009}%
  \BibitemOpen
  \bibfield  {author} {\bibinfo {author} {\bibnamefont {Stawarz}, \bibfnamefont
  {J.~E.}}, \bibinfo {author} {\bibnamefont {Smith}, \bibfnamefont {C.~W.}},
  \bibinfo {author} {\bibnamefont {Vasquez}, \bibfnamefont {B.~J.}}, \bibinfo
  {author} {\bibnamefont {Forman}, \bibfnamefont {M.~A.}}, \ and\ \bibinfo
  {author} {\bibnamefont {MacBride}, \bibfnamefont {B.~T.}},\ }\bibfield
  {title} {\enquote {\bibinfo {title} {The turbulent cascade and proton heating
  in the solar wind at 1 au},}\ }\href@noop {} {\bibfield  {journal} {\bibinfo
  {journal} {The Astrophysical Journal}\ }\textbf {\bibinfo {volume} {697}},\
  \bibinfo {pages} {1119} (\bibinfo {year} {2009})}\BibitemShut {NoStop}%
\bibitem [{\citenamefont {Stawarz}\ \emph {et~al.}(2010)\citenamefont
  {Stawarz}, \citenamefont {Smith}, \citenamefont {Vasquez}, \citenamefont
  {Forman},\ and\ \citenamefont {MacBride}}]{St2010}%
  \BibitemOpen
  \bibfield  {author} {\bibinfo {author} {\bibnamefont {Stawarz}, \bibfnamefont
  {J.~E.}}, \bibinfo {author} {\bibnamefont {Smith}, \bibfnamefont {C.~W.}},
  \bibinfo {author} {\bibnamefont {Vasquez}, \bibfnamefont {B.~J.}}, \bibinfo
  {author} {\bibnamefont {Forman}, \bibfnamefont {M.~A.}}, \ and\ \bibinfo
  {author} {\bibnamefont {MacBride}, \bibfnamefont {B.~T.}},\ }\bibfield
  {title} {\enquote {\bibinfo {title} {The turbulent cascade for high
  cross-helicity states at 1 au},}\ }\href@noop {} {\bibfield  {journal}
  {\bibinfo  {journal} {The Astrophysical Journal}\ }\textbf {\bibinfo {volume}
  {713}},\ \bibinfo {pages} {920} (\bibinfo {year} {2010})}\BibitemShut
  {NoStop}%
\bibitem [{\citenamefont {Stawarz}\ \emph {et~al.}(2011)\citenamefont
  {Stawarz}, \citenamefont {Vasquez}, \citenamefont {Smith}, \citenamefont
  {Forman},\ and\ \citenamefont {Klewicki}}]{St2011}%
  \BibitemOpen
  \bibfield  {author} {\bibinfo {author} {\bibnamefont {Stawarz}, \bibfnamefont
  {J.~E.}}, \bibinfo {author} {\bibnamefont {Vasquez}, \bibfnamefont {B.~J.}},
  \bibinfo {author} {\bibnamefont {Smith}, \bibfnamefont {C.~W.}}, \bibinfo
  {author} {\bibnamefont {Forman}, \bibfnamefont {M.~A.}}, \ and\ \bibinfo
  {author} {\bibnamefont {Klewicki}, \bibfnamefont {J.}},\ }\bibfield  {title}
  {\enquote {\bibinfo {title} {Third moments and the role of anisotropy from
  velocity shear in the solar wind},}\ }\href@noop {} {\bibfield  {journal}
  {\bibinfo  {journal} {The Astrophysical Journal}\ }\textbf {\bibinfo {volume}
  {736}},\ \bibinfo {pages} {44} (\bibinfo {year} {2011})}\BibitemShut
  {NoStop}%
\bibitem [{\citenamefont {Tu}\ and\ \citenamefont {Marsch}(1995)}]{Tu1995}%
  \BibitemOpen
  \bibfield  {author} {\bibinfo {author} {\bibnamefont {Tu}, \bibfnamefont
  {C.-Y.}}\ and\ \bibinfo {author} {\bibnamefont {Marsch}, \bibfnamefont
  {E.}},\ }\bibfield  {title} {\enquote {\bibinfo {title} {Mhd structures,
  waves and turbulence in the solar wind: Observations and theories},}\ }\href
  {\doibase 10.1007/BF00748891} {\bibfield  {journal} {\bibinfo  {journal}
  {Space Science Reviews}\ }\textbf {\bibinfo {volume} {73}},\ \bibinfo {pages}
  {1--210} (\bibinfo {year} {1995})}\BibitemShut {NoStop}%
\bibitem [{\citenamefont {Vasquez}\ \emph {et~al.}(2007)\citenamefont
  {Vasquez}, \citenamefont {Smith}, \citenamefont {Hamilton}, \citenamefont
  {MacBride},\ and\ \citenamefont {Leamon}}]{V2007}%
  \BibitemOpen
  \bibfield  {author} {\bibinfo {author} {\bibnamefont {Vasquez}, \bibfnamefont
  {B.~J.}}, \bibinfo {author} {\bibnamefont {Smith}, \bibfnamefont {C.~W.}},
  \bibinfo {author} {\bibnamefont {Hamilton}, \bibfnamefont {K.}}, \bibinfo
  {author} {\bibnamefont {MacBride}, \bibfnamefont {B.~T.}}, \ and\ \bibinfo
  {author} {\bibnamefont {Leamon}, \bibfnamefont {R.~J.}},\ }\bibfield  {title}
  {\enquote {\bibinfo {title} {Evaluation of the turbulent energy cascade rates
  from the upper inertial range in the solar wind at 1 au},}\ }\href {\doibase
  10.1029/2007JA012305} {\bibfield  {journal} {\bibinfo  {journal} {Journal of
  Geophysical Research: Space Physics}\ }\textbf {\bibinfo {volume} {112}}
  (\bibinfo {year} {2007}),\ 10.1029/2007JA012305},\ \Eprint
  {http://arxiv.org/abs/https://agupubs.onlinelibrary.wiley.com/doi/pdf/10.1029/2007JA012305}
  {https://agupubs.onlinelibrary.wiley.com/doi/pdf/10.1029/2007JA012305}
  \BibitemShut {NoStop}%
\bibitem [{\citenamefont {Verdini}\ and\ \citenamefont
  {Grappin}(2015)}]{Verdini_2015}%
  \BibitemOpen
  \bibfield  {author} {\bibinfo {author} {\bibnamefont {Verdini}, \bibfnamefont
  {A.}}\ and\ \bibinfo {author} {\bibnamefont {Grappin}, \bibfnamefont {R.}},\
  }\bibfield  {title} {\enquote {\bibinfo {title} {{IMPRINTS} {OF} {EXPANSION}
  {ON} {THE} {LOCAL} {ANISOTROPY} {OF} {SOLAR} {WIND} {TURBULENCE}},}\ }\href
  {\doibase 10.1088/2041-8205/808/2/l34} {\bibfield  {journal} {\bibinfo
  {journal} {The Astrophysical Journal}\ }\textbf {\bibinfo {volume} {808}},\
  \bibinfo {pages} {L34} (\bibinfo {year} {2015})}\BibitemShut {NoStop}%
\bibitem [{\citenamefont {Wan}\ \emph {et~al.}(2010)\citenamefont {Wan},
  \citenamefont {Servidio}, \citenamefont {Oughton},\ and\ \citenamefont
  {Matthaeus}}]{W2010}%
  \BibitemOpen
  \bibfield  {author} {\bibinfo {author} {\bibnamefont {Wan}, \bibfnamefont
  {M.}}, \bibinfo {author} {\bibnamefont {Servidio}, \bibfnamefont {S.}},
  \bibinfo {author} {\bibnamefont {Oughton}, \bibfnamefont {S.}}, \ and\
  \bibinfo {author} {\bibnamefont {Matthaeus}, \bibfnamefont {W.~H.}},\
  }\bibfield  {title} {\enquote {\bibinfo {title} {The third-order law for
  magnetohydrodynamic turbulence with shear: Numerical investigation},}\
  }\href@noop {} {\bibfield  {journal} {\bibinfo  {journal} {Phys. Plasmas}\
  }\textbf {\bibinfo {volume} {17}},\ \bibinfo {pages} {052307} (\bibinfo
  {year} {2010})}\BibitemShut {NoStop}%
\bibitem [{\citenamefont {Wang}\ \emph {et~al.}(2022)\citenamefont {Wang},
  \citenamefont {Chhiber}, \citenamefont {Adhikari}, \citenamefont {Yang},
  \citenamefont {Bandyopadhyay}, \citenamefont {Shay}, \citenamefont {Oughton},
  \citenamefont {Matthaeus},\ and\ \citenamefont {Cuesta}}]{Wang2022}%
  \BibitemOpen
  \bibfield  {author} {\bibinfo {author} {\bibnamefont {Wang}, \bibfnamefont
  {Y.}}, \bibinfo {author} {\bibnamefont {Chhiber}, \bibfnamefont {R.}},
  \bibinfo {author} {\bibnamefont {Adhikari}, \bibfnamefont {S.}}, \bibinfo
  {author} {\bibnamefont {Yang}, \bibfnamefont {Y.}}, \bibinfo {author}
  {\bibnamefont {Bandyopadhyay}, \bibfnamefont {R.}}, \bibinfo {author}
  {\bibnamefont {Shay}, \bibfnamefont {M.~A.}}, \bibinfo {author} {\bibnamefont
  {Oughton}, \bibfnamefont {S.}}, \bibinfo {author} {\bibnamefont {Matthaeus},
  \bibfnamefont {W.~H.}}, \ and\ \bibinfo {author} {\bibnamefont {Cuesta},
  \bibfnamefont {M.~E.}},\ }\bibfield  {title} {\enquote {\bibinfo {title}
  {Strategies for determining the cascade rate in {MHD} turbulence: Isotropy,
  anisotropy, and spacecraft sampling},}\ }\href {\doibase
  10.3847/1538-4357/ac8f90} {\bibfield  {journal} {\bibinfo  {journal} {The
  Astrophysical Journal}\ }\textbf {\bibinfo {volume} {937}},\ \bibinfo {pages}
  {76} (\bibinfo {year} {2022})}\BibitemShut {NoStop}%
\bibitem [{\citenamefont {Weygand}\ \emph {et~al.}(2007)\citenamefont
  {Weygand}, \citenamefont {Matthaeus}, \citenamefont {Dasso}, \citenamefont
  {Kivelson},\ and\ \citenamefont {Walker}}]{WEY2007}%
  \BibitemOpen
  \bibfield  {author} {\bibinfo {author} {\bibnamefont {Weygand}, \bibfnamefont
  {J.~M.}}, \bibinfo {author} {\bibnamefont {Matthaeus}, \bibfnamefont
  {W.~H.}}, \bibinfo {author} {\bibnamefont {Dasso}, \bibfnamefont {S.}},
  \bibinfo {author} {\bibnamefont {Kivelson}, \bibfnamefont {M.~G.}}, \ and\
  \bibinfo {author} {\bibnamefont {Walker}, \bibfnamefont {R.~J.}},\ }\bibfield
   {title} {\enquote {\bibinfo {title} {Taylor scale and effective magnetic
  reynolds number determination from plasma sheet and solar wind magnetic field
  fluctuations},}\ }\href@noop {} {\bibfield  {journal} {\bibinfo  {journal}
  {J. Geophys. Res.: Space Phys.}\ }\textbf {\bibinfo {volume} {112}},\
  \bibinfo {pages} {A10} (\bibinfo {year} {2007})}\BibitemShut {NoStop}%
\bibitem [{\citenamefont {de~Wit}\ \emph {et~al.}(2020)\citenamefont {de~Wit},
  \citenamefont {Krasnoselskikh}, \citenamefont {Bale}, \citenamefont
  {Bonnell}, \citenamefont {Bowen}, \citenamefont {Chen}, \citenamefont
  {Froment}, \citenamefont {Goetz}, \citenamefont {Harvey}, \citenamefont
  {Jagarlamudi}, \citenamefont {Larosa}, \citenamefont {MacDowall},
  \citenamefont {Malaspina}, \citenamefont {Matthaeus}, \citenamefont {Pulupa},
  \citenamefont {Velli},\ and\ \citenamefont {Whittlesey}}]{Dudok2020}%
  \BibitemOpen
  \bibfield  {author} {\bibinfo {author} {\bibnamefont {de~Wit}, \bibfnamefont
  {T.~D.}}, \bibinfo {author} {\bibnamefont {Krasnoselskikh}, \bibfnamefont
  {V.~V.}}, \bibinfo {author} {\bibnamefont {Bale}, \bibfnamefont {S.~D.}},
  \bibinfo {author} {\bibnamefont {Bonnell}, \bibfnamefont {J.~W.}}, \bibinfo
  {author} {\bibnamefont {Bowen}, \bibfnamefont {T.~A.}}, \bibinfo {author}
  {\bibnamefont {Chen}, \bibfnamefont {C.~H.~K.}}, \bibinfo {author}
  {\bibnamefont {Froment}, \bibfnamefont {C.}}, \bibinfo {author} {\bibnamefont
  {Goetz}, \bibfnamefont {K.}}, \bibinfo {author} {\bibnamefont {Harvey},
  \bibfnamefont {P.~R.}}, \bibinfo {author} {\bibnamefont {Jagarlamudi},
  \bibfnamefont {V.~K.}}, \bibinfo {author} {\bibnamefont {Larosa},
  \bibfnamefont {A.}}, \bibinfo {author} {\bibnamefont {MacDowall},
  \bibfnamefont {R.~J.}}, \bibinfo {author} {\bibnamefont {Malaspina},
  \bibfnamefont {D.~M.}}, \bibinfo {author} {\bibnamefont {Matthaeus},
  \bibfnamefont {W.~H.}}, \bibinfo {author} {\bibnamefont {Pulupa},
  \bibfnamefont {M.}}, \bibinfo {author} {\bibnamefont {Velli}, \bibfnamefont
  {M.}}, \ and\ \bibinfo {author} {\bibnamefont {Whittlesey}, \bibfnamefont
  {P.~L.}},\ }\bibfield  {title} {\enquote {\bibinfo {title} {Switchbacks in
  the near-sun magnetic field: Long memory and impact on the turbulence
  cascade},}\ }\href {\doibase 10.3847/1538-4365/ab5853} {\bibfield  {journal}
  {\bibinfo  {journal} {The Astrophysical Journal Supplement Series}\ }\textbf
  {\bibinfo {volume} {246}},\ \bibinfo {pages} {39} (\bibinfo {year}
  {2020})}\BibitemShut {NoStop}%
\end{thebibliography}%
